\documentclass[journal,comsoc]{IEEEtran}
\usepackage{amsmath,amssymb,amsfonts,chemarrow,balance}
\usepackage{hyperref}
\usepackage{amsmath} 
\usepackage{graphicx}
\usepackage{subfigure}
\usepackage{url}
\usepackage{mathenv}
\usepackage{color}
\usepackage{breqn}
 \usepackage{algpseudocode}
\newcommand{\BfPara}[1]{{\noindent\bf#1.}\xspace}
\usepackage{multirow}
\usepackage{float}
\usepackage{threeparttable}
\usepackage[inline,shortlabels]{enumitem}
\usepackage{wasysym}
\usepackage{etoolbox}

\usepackage[T1]{fontenc}
\usepackage[utf8]{inputenc}
\usepackage{environ}
\usepackage{tikz}
\usetikzlibrary{arrows}
\usetikzlibrary{calc,matrix}
\renewcommand{\baselinestretch}{0.975} 
\usepackage{xspace}
\usepackage{setspace}
\newcommand{\note}[1]{}

\newcommand{\bc}{{Bitcoin}\xspace}
\newcommand{\bcc}{{Blockchain}\xspace}

\newcommand{\etc}{{etc.}\xspace}
\newcommand{\eg}{{\em e.g.}\xspace}
\newcommand{\ie}{{\em i.e.,}\xspace}
\newcommand{\cc}{{cryptocurrency}\xspace}

\newcommand{\cj}{{cryptojacking}\xspace}
\newcommand{\Cj}{{Cryptojacking}\xspace}
\newcommand{\ddos}{{DDoS}\xspace}

\newcommand{\etal}{{\em et al.}\xspace}

\usepackage{cite}
\usepackage{hyperref}
\definecolor{linkcolour}{rgb}{0,0.2,0.6}
\definecolor{xgreen}{rgb}{0.2,0.6,0.0}
\definecolor{xred}{rgb}{0.7,0.1,0.0}
\def\equationautorefname~#1\null{(#1)\null}

\usepackage{listings}
\colorlet{punct}{red!60!black}
\definecolor{background}{HTML}{ffffff }
\definecolor{delim}{RGB}{20,105,176}
\colorlet{numb}{magenta!60!black}

\definecolor{light-gray}{gray}{0.95}
\definecolor{darkgray}{rgb}{0.4, 0.4, 0.4}
\definecolor{editorGray}{rgb}{0.95, 0.95, 0.95}
\definecolor{editorOcher}{rgb}{1, 0.5, 0} 
\definecolor{editorGreen}{rgb}{0, 0.5, 0} 
\definecolor{orange}{rgb}{1,0.45,0.13}      
\definecolor{olive}{rgb}{0.17,0.59,0.20}
\definecolor{brown}{rgb}{0.69,0.31,0.31}
\definecolor{purple}{rgb}{0.38,0.18,0.81}
\definecolor{lightblue}{rgb}{0.1,0.57,0.7}
\definecolor{lightred}{rgb}{1,0.4,0.5}
\usepackage{upquote}
\usepackage{listings}

\definecolor{pblue}{rgb}{0.13,0.13,1}
\definecolor{pgreen}{rgb}{0,0.5,0}
\definecolor{pred}{rgb}{0.9,0,0}
\definecolor{pgrey}{rgb}{0.46,0.45,0.48}

\makeatletter
\let\matamp=&
\catcode`\&=13
\makeatletter
\def&{\iftikz@is@matrix
  \pgfmatrixnextcell
  \else
  \matamp
  \fi}
\makeatother

\newcounter{lines}
\def\endlr{\stepcounter{lines}\\}

\newcounter{vtml}
\newif\ifvtimelinetitle
\newif\ifvtimebottomline
\tikzset{description/.style={
  column 2/.append style={#1}
 },
 timeline color/.store in=\vtmlcolor,
 timeline color=red!80!black,
 timeline color st/.style={fill=\vtmlcolor,draw=\vtmlcolor},
 use timeline header/.is if=vtimelinetitle,
 use timeline header=false,
 add bottom line/.is if=vtimebottomline,
 add bottom line=false,
 timeline title/.store in=\vtimelinetitle,
 timeline title={},
 line offset/.store in=\lineoffset,
 line offset=4pt,
}

\NewEnviron{vtimeline}[1][]{%
\setcounter{lines}{1}%
\stepcounter{vtml}%
\begin{tikzpicture}[column 1/.style={anchor=east},
 column 2/.style={anchor=west},
 text depth=0pt,text height=1ex,
 row sep=1ex,
 column sep=1em,
 #1
]
\matrix(vtimeline\thevtml)[matrix of nodes]{\BODY};
\pgfmathtruncatemacro\endmtx{\thelines-1}
\path[timeline color st] 
($(vtimeline\thevtml-1-1.north east)!0.5!(vtimeline\thevtml-1-2.north west)$)--
($(vtimeline\thevtml-\endmtx-1.south east)!0.5!(vtimeline\thevtml-\endmtx-2.south west)$);
\foreach \x in {1,...,\endmtx}{
 \node[circle,timeline color st, inner sep=0.15pt, draw=white, thick] 
 (vtimeline\thevtml-c-\x) at 
 ($(vtimeline\thevtml-\x-1.east)!0.5!(vtimeline\thevtml-\x-2.west)$){};
 \draw[timeline color st](vtimeline\thevtml-c-\x.west)--++(-3pt,0);
 }
 \ifvtimelinetitle%
  \draw[timeline color st]([yshift=\lineoffset]vtimeline\thevtml.north west)--
  ([yshift=\lineoffset]vtimeline\thevtml.north east);
  \node[anchor=west,yshift=16pt,font=\large]
   at (vtimeline\thevtml-1-1.north west) 
   {\textsc{Timeline \thevtml}: \textit{\vtimelinetitle}};
 \else%
  \relax%
 \fi%
 \ifvtimebottomline%
   \draw[timeline color st]([yshift=-\lineoffset]vtimeline\thevtml.south west)--
  ([yshift=-\lineoffset]vtimeline\thevtml.south east);
 \else%
   \relax%
 \fi%
\end{tikzpicture}
}

\begin{document}

\title{Exploring the Attack Surface of Blockchain: \\ A Systematic Overview}

\author{Muhammad Saad\thanks{M. Saad, J. Spaulding, and A. Mohaisen are with the Department of Computer Science at the University of Central Florida, Florida 32816, USA. L. Njilla is with the Air Force Research Laboratory, Rome, NY, USA. C. Kamhoua is with the Army Research Laboratory, MD, USA. S. Shetty is with the Old Dominion University, VA, USA. D. Nyang is with INHA University, Incheon, South Korea. The work of D. Nyang was done while visiting the University of Central Florida.}, Jeffrey Spaulding, Laurent Njilla, Charles Kamhoua, \\Sachin Shetty, DaeHun Nyang, and Aziz Mohaisen}

\maketitle

\begin{abstract}
In this paper, we systematically explore the attack surface of the Blockchain technology, with an emphasis on public Blockchains. Towards this goal, we attribute attack viability in the attack surface to 1) the Blockchain cryptographic constructs, 2) the distributed architecture of the systems using Blockchain, and 3) the Blockchain application context. To each of those contributing factors, we outline several attacks, including selfish mining, the 51\% attack, Domain Name System (DNS) attacks, distributed denial-of-service (DDoS) attacks, consensus delay (due to selfish behavior or distributed denial-of-service attacks), Blockchain forks, orphaned and stale blocks, block ingestion, wallet thefts, smart contract attacks, and privacy attacks. We also explore the causal relationships between these attacks to demonstrate how various attack vectors are connected to one another. A secondary contribution of this work is outlining effective defense measures taken by the Blockchain technology or proposed by researchers to mitigate the effects of these attacks and patch associated vulnerabilities. 

\end{abstract}
\begin{IEEEkeywords} Blockchain; Security; Attack Surface; Applications; Peer-to-Peer Systems \end{IEEEkeywords}
\IEEEpeerreviewmaketitle

\section{Introduction}\label{sec:introduction}

Blockchain technology is being explored in many innovative applications, such as cryptocurrencies~\cite{MauriCD18,DanezisM16,BonneauMCNKF15}, smart contracts~\cite{kosba2016hawk,BhargavanDFGGKK16}, communication systems~\cite{SharmaRP18,FanRWLY18}, health care~
\cite{GuoSZZ18,Rakic18}, Internet of Things~\cite{JesusCAR18,SharmaSJP17}, financial systems~\cite{HyvarinenRF17,HolotiukPM17}, censorship resistance~\cite{HeilmanBG16}, electronic voting~\cite{DagherMMM18,HardwickA18}, and distributed provenance~\cite{EljazzarAKE18,BaruffaldiS18,FotiouP16}, among others. Using Blockchain's transparent and fully distributed peer-to-peer architecture, these applications benefit from an append-only model in which ``transactions'' accepted in the Blockchain cannot be modified \cite{ZhangJ18a,Mettler16,BaruffaldiS18}. The transparency of the Blockchain enables storing publicly verifiable and undeniable records~\cite{zyskind2015decentralizing}. Furthermore, the Blockchain's peer-to-peer system provides verifiable ledger maintenance without a centralized authority, thus addressing the single point-of-failure and single point-of-trust~\cite{back2014enabling}. For instance, \bc (a popular \cc using Blockchain technology) takes advantage of the aforementioned properties, making it easy to verify the history of financial transactions~\cite{nakamoto2008bitcoin,RuffingMK17}.

\begin{table*}[t]
\small
\centering
\caption{Attack vectors related to the attack class in Blockchain systems. We also show, by referencing to the prior work, how each attack affects the entities involved with Blockchain systems. For instance, Orphaned blocks affect the Blockchain, the miners, and the mining pools. }
\begin{center}

\begin{tabular}{|c|l|c|c|c|c|c|c|}\hline
\multirow{1}{*}{{}} & \multirow{1}{*}{{Attacks}} & \multirow{1}{*}{{Blockchain}}     & \multirow{1}{*}{{Miners}} & \multirow{1}{*}{{Mining Pools}} & \multirow{1}{*}{{Exchanges}} & \multirow{1}{*}{{Application}} & \multirow{1}{*}{{Users}} \\ \cline{1-8}

\multirow{2}{*}{{Blockchain Structure}} & \multirow{1}{*}{Forks \cite{eyal2015miner}} &\checkmark  & & & & &  \\  \cline{2-8}
    & \multirow{1}{*}{Orphaned blocks \cite{decker2013information}} & \checkmark & \checkmark & \checkmark & & &  \\ \cline{1-8}

\multirow{9}{*}{{Peer-to-Peer System}} & \multirow{1}{*}{DNS hijacks \cite{developer}} & & \checkmark  & \checkmark & \checkmark & & \checkmark \\ \cline{2-8}
    & \multirow{1}{*}{BGP hijacks \cite{apostolaki2017hijacking} } & &\checkmark  &\checkmark & & & \checkmark  \\ \cline{2-8}
    & \multirow{1}{*}{Eclipse attack \cite{MarcusHG18} } & & \checkmark & & & & \checkmark  \\ \cline{2-8}
    & \multirow{1}{*}{Majority attack \cite{bastiaan2015preventing} } &\checkmark & \checkmark  & & &\checkmark &   \\ \cline{2-8}
    & \multirow{1}{*}{Selfish mining \cite{LeelavimolsilpTS18}} & \checkmark & \checkmark  & \checkmark & & &  \\ \cline{2-8}
    & \multirow{1}{*}{DDoS attacks \cite{saad2018poster}} & \checkmark & \checkmark  &\checkmark & & &  \\ \cline{2-8}
    & \multirow{1}{*}{Consensus Delay \cite{eyal2016bitcoin}} & & \checkmark & \checkmark & & &\checkmark  \\ \cline{2-8}
    & \multirow{1}{*}{Block Withholding \cite{eyal2016bitcoin}} & & \checkmark  & \checkmark & & &  \\ \cline{2-8}
    & \multirow{1}{*}{Timejacking attacks \cite{vyas2014security} } & &\checkmark & \checkmark & & \checkmark &  \\ \cline{2-8}
    & \multirow{1}{*}{Finney attacks \cite{finney-attack} } & & \checkmark & \checkmark & & &\checkmark   \\ \cline{1-8}

\multirow{10}{*}{{Blockchain Application}} & \multirow{1}{*}{Blockchain Ingestion \cite{fleder2015bitcoin}} & \checkmark &  & & & &   \\ \cline{2-8}
    & \multirow{1}{*}{Wallet theft \cite{bamert2014bluewallet} } & &  &  &\checkmark & \checkmark &\checkmark  \\ \cline{2-8}
    & \multirow{1}{*}{Double-spending \cite{doublespending}} & \checkmark &  & &  & &\checkmark    \\ \cline{2-8}
    & \multirow{1}{*}{Cryptojacking \cite{TahirHDAGZCB17} }  & &  & & & \checkmark &\checkmark  \\ \cline{2-8}
        & \multirow{1}{*}{Smart contract DoS \cite{ethereum-safety}} & \checkmark &  & & &\checkmark & \checkmark \\ \cline{2-8}
    & \multirow{1}{*}{$\approx$ Reentracy attacks \cite{grincalaitis_2017}} & &  & & &\checkmark & \checkmark  \\ \cline{2-8}
    & \multirow{1}{*}{$\approx$ Overflow attacks \cite{grincalaitis_2017}} & &  & & & \checkmark & \checkmark  \\ \cline{2-8}
    & \multirow{1}{*}{$\approx$ Replay attacks \cite{ethereum-safety}} & \checkmark &  & \checkmark & & \checkmark & \checkmark  \\ \cline{2-8}
    & \multirow{1}{*}{$\approx$ Short address attacks \cite{grincalaitis_2017}} & &  & & &\checkmark &  \\ \cline{2-8}
    & \multirow{1}{*}{$\approx$ Balance attacks \cite{ethereum-safety}} & &  & & &\checkmark & \checkmark \\ \cline{1-8}

\end{tabular}
\end{center}
\label{tab:multicol}
\end{table*}

\begin{table*}[t]

\centering
\caption{Implications of each attack on the Blockchain system in the light of the prior work. For instance, forks can lead to chain splitting and revenue loss. As a result of a fork, one among the candidate chains is selected by the network while the others are invalidated. This leads to invalidation of transaction and revenue loss to miners. }

\begin{center}
\scalebox{0.85}{
\begin{tabular}{|c|l|c|c|c|c|c|c|c|}\hline
\multirow{1}{*}{{}} & \multirow{1}{*}{{Attacks}} & \multirow{1}{*}{{ Chain Splitting }} & \multirow{1}{*}{{ Revenue Loss}} & \multirow{1}{*}{{Partitioning}} & \multirow{1}{*}{{Malicious Mining}} & \multirow{1}{*}{{Delay}} & \multirow{1}{*}{{Info Loss}} & \multirow{1}{*}{{Theft}}\\ \cline{1-9}
  
\multirow{2}{*}{{\shortstack[l]{Blockchain\\ Splitting}}} & \multirow{1}{*}{Forks \cite{eyal2015miner}} &\checkmark & \checkmark & & & & & \\  \cline{2-9}
& \multirow{1}{*}{Orphaned Blocks \cite{decker2013information}} &  & \checkmark &  & & & & \\ \cline{1-9}

\multirow{9}{*}{{\shortstack[l]{P2P\\ System}}} & \multirow{1}{*}{DNS hijacks \cite{developer}} & & \checkmark    & \checkmark  &   & &  & \checkmark \\ \cline{2-9}
& \multirow{1}{*}{BGP hijacks \cite{apostolaki2017hijacking} } & & \checkmark  & \checkmark & & &   & \checkmark \\ \cline{2-9}
& \multirow{1}{*}{Eclipse attacks \cite{MarcusHG18}} & &   &  \checkmark& & &   & \\ \cline{2-9}
& \multirow{1}{*}{Majority attacks \cite{bastiaan2015preventing} } & \checkmark & \checkmark   & &\checkmark &  & &  \\ \cline{2-9}
& \multirow{1}{*}{Selfish mining \cite{LeelavimolsilpTS18}} &   & \checkmark   &   &\checkmark & & & \\ \cline{2-9}
& \multirow{1}{*}{DDoS attacks \cite{saad2018poster}} &   &    &  & \checkmark & & & \checkmark \\ \cline{2-9}
& \multirow{1}{*}{Consensus Delay \cite{eyal2016bitcoin}} & &    &    & & \checkmark & \checkmark  & \\ \cline{2-9}
& \multirow{1}{*}{Block Withholding \cite{eyal2016bitcoin} } & & \checkmark     &    & \checkmark & & & \\ \cline{2-9}
& \multirow{1}{*}{Timejacking attacks \cite{vyas2014security} } & \checkmark & \checkmark  &    & \checkmark &  \checkmark & & \\ \cline{2-9}
& \multirow{1}{*}{Finney attacks \cite{finney-attack} } & &  \checkmark  &    & & &   &  \\ \cline{1-9}

\multirow{10}{*}{{\shortstack[l]{Blockchain\\ Application}}} & \multirow{1}{*}{Blockchain Ingestion \cite{fleder2015bitcoin}} &    &  & & & &\checkmark &  \\ \cline{2-9}
& \multirow{1}{*}{Wallet theft \cite{bamert2014bluewallet}} & & \checkmark &  &   &    &   & \checkmark \\ \cline{2-9}
& \multirow{1}{*}{Double-spending \cite{doublespending}} &    &  & &  & &    &  \\ \cline{2-9}
& \multirow{1}{*}{Cryptojacking \cite{TahirHDAGZCB17} }  & \checkmark &  & &\checkmark &   &  & \checkmark \\ \cline{2-9}
& \multirow{1}{*}{Smart contract DoS \cite{ethereum-safety}} & & \checkmark & & &\checkmark & & \checkmark \\ \cline{2-9}
& \multirow{1}{*}{$\approx$ Reentracy attacks \cite{grincalaitis_2017}} & & \checkmark  & & & & & \checkmark \\ \cline{2-9}
& \multirow{1}{*}{$\approx$ Overflow attacks \cite{grincalaitis_2017}} & &  & & & & & \checkmark \\ \cline{2-9}
& \multirow{1}{*}{$\approx$ Replay attacks \cite{ethereum-safety}} & & \checkmark & & & & \checkmark &  \\ \cline{2-9}
& \multirow{1}{*}{$\approx$ Short address attacks \cite{grincalaitis_2017}} & &\checkmark  & & & & & \checkmark \\ \cline{2-9}
& \multirow{1}{*}{$\approx$ Balance attacks \cite{ethereum-safety}} & & \checkmark  & & & & & \checkmark \\ \cline{1-9}

\end{tabular}}
\end{center}
\label{tab:multicol_2}
\end{table*}

Despite the functional features that Blockchain brings to the design space of these applications~\cite{underwood2016blockchain}, recent reports have highlighted the security risks associated with this technology~\cite{li2017survey,JesusCAR18,lin2017survey,AtzeiBC17,KhalilovL18}. For instance, in June 2016 an unknown attacker managed to drain \$50 million USD from ``The DAO'', a decentralized autonomous organization that operates on Blockchain-based smart contracts, or pre-programmed rules that govern the organization \cite{DAO}. In August 2016, bitcoins worth \$72 million USD were stolen from the exchange platform Bitfinex in Hong Kong \cite{baldwin_2016}. In June 2017, Bitfinex also experienced a distributed denial-of-service (DDoS) attack that led to its temporary suspension. Several exchanges of Bitcoin and Ethereum (a Blockchain-based distributed computing platform) have also suffered from DDoS attacks and DNS attacks frequently, hampering the service availability to the users. 

Often times these attacks are launched on blockchain-applications due to their popularity or the capital involved in their system. For instance, with \bc, such attacks can cause devaluation of the \cc, loss of mining rewards, or even closure of \cc exchanges~\cite{apostolaki2017hijacking}. \bc's Blockchain is also targeted with dust or spam transactions to delay the processing of legitimate transactions. In May, August, and November 2017, memory pools of Bitcoin were flooded with dust transactions to create stalls and delays in transaction verification, and to increase \bc mining fee \cite{saad2018poster}. The transaction stall in November 2017, for example, resulted in a payment delay of \$700 million USD worth bitcoins \cite{cryptocoinsnews_2017}. Often the intent of such attacks is to motivate the Bitcoin users to move to other cryptocurrencies with faster transaction processing time. 

Due to a publicly verifiable nature, Blockchain-based cryptocurrencies are vulnerable to several fraudulent activities. Mt. Gox, a Bitcoin currency exchange in Japan, was attacked by two malicious users who stole \$460 million USD worth of bitcoins \cite{McMillan14}. The attackers gathered useful information from Bitcoin's Blockchain and engineered a fake transaction ripple to increase the market price. Due to such activities, Mt. Gox suffered a heavy loss and eventually became bankrupt.

In May and June 2018, five Blockchain-based cryptocurrencies; namely, Monacoin, Bitcoin Gold, Zencash, Verge, and Litecoin Cash, were targeted by a 51\% attack~\cite{source}, leading to a loss of \$5 million USD. The attackers in each \cc were able to gain more than 51\% of the networks' hash rate which was used to rearrange transactions and prevent other miners from computing blocks. As a result, they were able to gain control over the Blockchain and perform double-spending on valuable transactions~\cite{Perez-SolaDNH17,KarameAC12}.

The security of \bcc systems is important for their acceptability by potential users~\cite{pilkington201611}. For example, investors take the security of \bc into account when studying the risks associated with their investments and use of this technology. Understanding the threats associated with \bcc systems in general is a first step towards realizing the potential of applications built on it. To this end, this work is dedicated to an in-depth look at the attack surface of \bcc. 

We envision that \bcc will be used in many applications, and we report on the attacks that could compromise those applications. Namely, the taxonomy of Blockchain attacks in this paper is classified into three broad categories: 1) attacks associated with the mathematical techniques used for creating the ledger (\eg, Blockchain forks, stale blocks, orphaned blocks, \etc), 2) attacks associated with the peer-to-peer architecture used in the \bcc system, (\eg, selfish mining, the 51\% attack, consensus delay, \ddos attack, Domain Name System (DNS) attacks, Fork After Withholding (FAW) Attacks, \etc ), and 3) attacks associated with the application context that uses the \bcc technology (\eg, Blockchain ingestion, double-spending, wallet theft \cite{DmitrienkoNY17}, \etc ). In this paper, we mainly focus on the attack surface of public and permissionless Blockchains. Public Blockchains are suitable for applications that provide open access to system resources while preserving user anonymity. These attributes are well suited for a system that has a weak trust model and high provenance assurance requirements. The weak trust model results from an application's tolerance for adversaries who can game the system while staying anonymous. On the other hand, high provenance means that anyone can access the publicly available resources to transparently audit data. For instance, in Bitcoin and Ethereum, any user can join the network by running an Ethereum software client on their machine and participating in transaction processing. Since the Blockchain is public, anyone outside the system can validate the authenticity of transactions and blocks. Therefore, public Blockchains remain a dominant component among Blockchain applications as shown by the popularity of Bitcoin and Ethereum. On the other hand, the weak trust model exposes public Blockchains to a wide variety of attacks, allowing adversaries to easily compromise the system~\cite{ContiELR17}. Therefore, while the public Blockchains are useful for an open access system, they are not suitable for closed environments where the weak trust model creates attack opportunities.

To address the shortcomings of public Blockchains and reduce the attack opportunities, private and permissioned Blockchains are now used for various applications~\cite{DinhW0LOT17}. In private Blockchains, the access to system resources is restricted to a chosen set of peers~\cite{BarallaIMTM18,AhmadSBM18}. These peers are screened prior to their induction in the application. Since the information about peers is known, their identities can be tied (or attributed) to their behavior in order to prevent attacks. Although private Blockchains still act as agents of trust in permissioned settings, they are not significantly exposed to adversarial attacks due to a stronger trust model. Since the aim of this work is to explore and understand the attack surface of Blockchains, it is natural to focus more on the public Blockchains. However, wherever necessary, we will also discuss the security and performance of private Blockchains as well.

\BfPara{Contributions} In summary, we make the following contributions int this paper. 
\begin{enumerate*}[(1)]
\item We survey the possible attacks related to the design constructs of Blockchains, the peer-to-peer architecture, and the application-oriented use of Blockchains.
\item We explore the origins of these attacks and the ways in which they affect Blockchain applications and their users. 
\item We also show the relationship between a sequence of attacks to outline how one attack can facilitate the possibility of other attacks. Understanding these links can help devise a common cure that can fix multiple problems at the same time.
\item Building on top of the prior work~\cite{ContiELR17,lin2017survey,AtzeiMT17}, for each attack class, we also explore the possible defense strategies that have been proposed to harden the security of Blockchains. Since many attacks related to a specific class have a common defense or remedy, while others remain as open problems, we discuss combined countermeasures for each class.  Moreover, by highlighting the lessons learned, we also provide future research directions towards a more systematic treatment of the Blockchain attack surface.
\item In~\autoref{tab:multicol} and \autoref{tab:multicol_2}, we provide an overview of the Blockchain attack surface. We ascribe various attacks to attack classes with their implications.
\end{enumerate*}

\BfPara{Organization} The rest of the paper is organized as follows. In \autoref{sec:Motivation}, we provide the motivation of this work. In~\autoref{sec:Blockchain}, we give an overview of Blockchain its operations.  In~\autoref{sec:bcstructure}, we review the design constructs of \bcc that enable various attacks, such as Blockchain forks, stale and orphaned blocks. In~\autoref{sec:p2p}, we look into the features of distributed networks that create possibilities for the 51\% attack, DNS attacks, DDoS attacks, consensus delays, \etc We further describe the aspects of peer-to-peer architecture that enable the possibility of their potential misuse in \bcc applications.  In~\autoref{sec:application}, we outline the application-specific vulnerabilities found in \bcc and assess the threats that they face. That is followed by discussion and open directions in \autoref{sec:disc}, and the concluding remarks in~\autoref{sec:conclusion}.

\begin{figure*}[htb]
\begin{center}
\includegraphics[ width=1\textwidth]{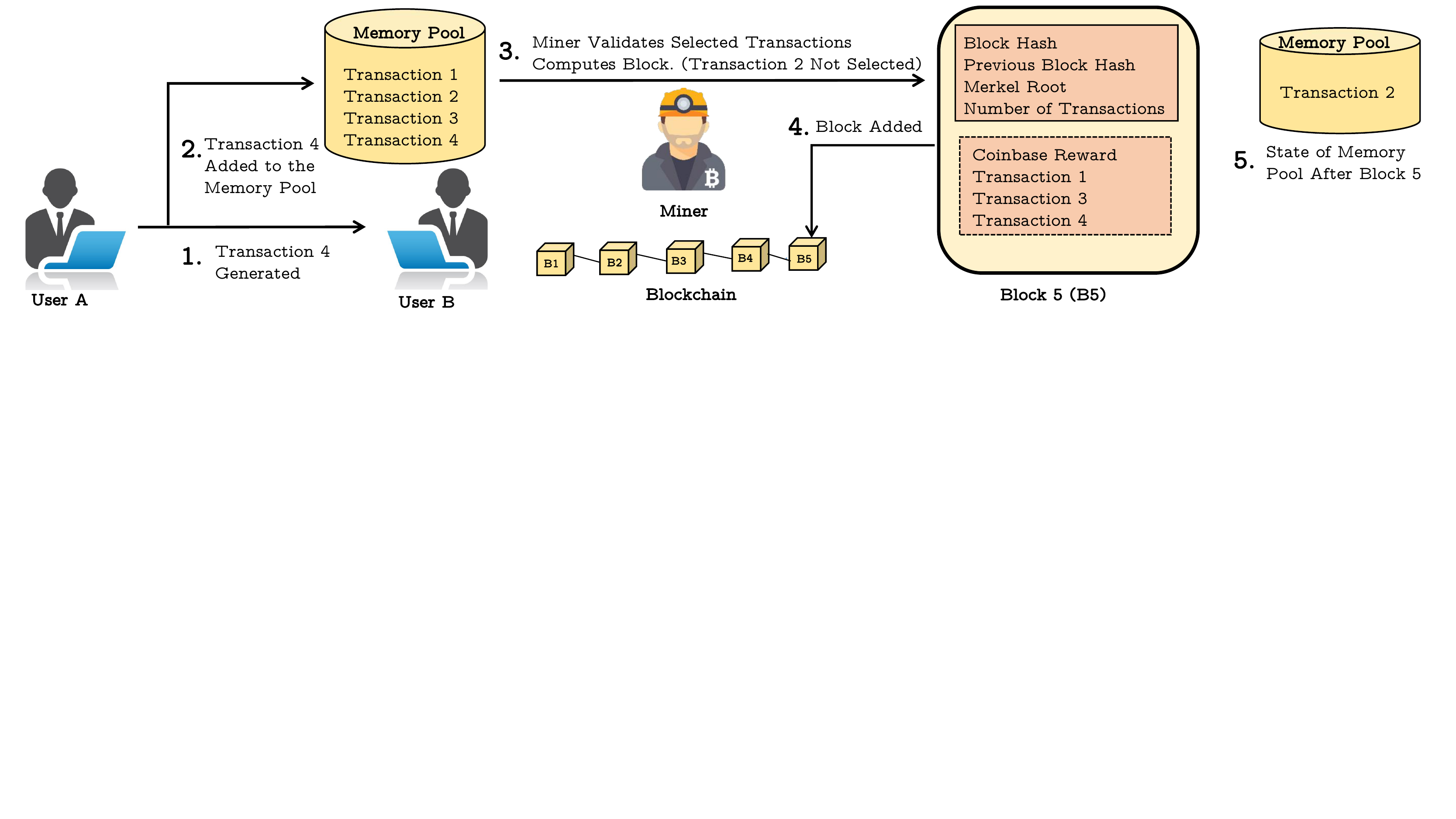}
\vspace{-62mm}
\caption{Transaction life-cycle in a PoW-based cryptocurrency. User A generates a transaction for user B. The transaction is stored in the memory pool along with other unconfirmed transactions. Miner validates transactions from memory pool, and computes a block. A valid block is added to the Blockchain. }
\label{fig:overview}
\end{center}
\end{figure*}

\section{Motivation and Target Audience}\label{sec:Motivation}
The motivation of this work is to derive attention towards the security vulnerabilities of Blockchain systems via a systematic and comprehensive study. Recently, Blockchain technology has gained significant attention and its applications are being explored in various domains~\cite{OuaddahKO16,ChristidisD16}. Blockchains are capable of augmenting trust and provide provenance in distributed systems. While acknowledging their merits~\cite{MillerBKM17,LindNEKPS18,LundbaekDH16}, we argue that it is important to understand their shortcomings, particularly related to security, as evident by the large security surface. To that end, our work is an effort to highlight potential vulnerabilities in Blockchains, with an emphasis on popular public Blockchain applications. We systematically analyze various attack vectors and study their relationships. Alongside, we also survey countermeasures and defenses to the various attack surface elements, and provide future research directions.

Since various research and technology sections are interested in using Blockchains, it is intuitive to explore a deeper understanding of Blockchains' attack surface to establish foundations for their security. For instance, using public Blockchains in the financial sector may prevent fraud and data tampering, by the simply utilizing Blockains' properties, although that also may expose sensitive information of financial transactions to adversaries. Similarly, organizations that are exploring Blockchain-based smart systems \cite{eyal2016bitcoin,MinaeiMK18}, while might benefit immensely in addressing functional requirements, need to be aware of the programming languages' constraints and shortcomings, as well as compilation bugs that may lead to data breach and critical assets loss. For this research-driven efforts, we believe our work has the potential to offer future directions toward designing more secure and robust Blockchain solutions that may overcome some of those challenges as outlined in the rest of this survey. Some of these challenges include constructing new consensus algorithms that are secure, scalable, and energy efficient~\cite{BissasLOAH16}. Additionally, they must also have the capability to prevent race conditions that lead to attacks such as selfish mining, double-spending, majority attacks, and orphaned blocks~\cite{GoldbergH18,RitzZ18}. To facilitate the process of addressing those challenges, we supplement our work by surveying the existing countermeasures proposed in the literature. These countermeasures can be used as building blocks for more secure and robust solutions.

In summary, the target audience of this work include both academics, who are interested in understanding the attack landscape of Blockchains, as well as practitioners, who might be interested in understanding the existing solutions to those attacks, to utilize as building blocks, and both benefiting from  a systematic analysis of the Blockchain attack surface.

\section{Overview of Blockchain and its Operations}\label{sec:Blockchain}
Conceptually, \bcc can be viewed as a repository of data that is tamper-evident due to its replication over all nodes in a peer-to-peer system. Transactions represent the events that drive the Blockchain application (\eg, in cryptocurrencies, tokens are the transactions exchanged among the users). Blockchain applications use various consensus algorithms for trust among peers over the state of the ledger. Moreover, the consensus algorithms ensure a consistent and transparent view of the Blockchain, thereby resolving conflicts and forks. This is, no block is added to the Blockchain, until it fulfills the conditions outlined by the consensus algorithm. Moreover, each algorithm has unique functional and operational properties that drive the consensus over the Blockchain.

While the consensus algorithms in Blockchains may vary, however, the Blockchain data structure and its network architecture remain consistent across all applications. For instance, in the two popular Blockchain applications, namely Bitcoin and Peercoin, the consensus algorithms are proof-of-work and proof-of-stake, respectively. Although they different in the way the consensus is conducted, the two applications have the same Blockchain data structure in which the chain progresses in an append-only model and each block is linked to the previous block through a one-way hash function. In both cryptocurrencies, the system replicas are connected in a peer-to-peer model, maintaining a single copy of the ledger. In the following, we briefly discuss the popular consensus algorithms along with the fundamental cryptographic primitives that are used in Blockchains.

\subsection{Consensus Algorithms}\label{sec:cons}
Some of the notable consensus algorithms used in Blockchains include proof-of-work (PoW), proof-of-stake (PoS), proof-of-activity (PoA), proof-of-capacity (PoC), proof-of-burn (PoB), proof-of-knowledge (PoK), and the practical Byzantine fault tolerance (PBFT)~\cite{BanoSBAMMD17,bellare1993random,juels1999client,coindesk_proofs,SaadM18}. The most popular consensus algorithm widely used in Blockchains is PoW, followed by PoS and PBFT. We discuss them in the following.

\BfPara{Proof-of-Work} In PoW Blockchains, peers in the network try to solve a computationally expensive mathematical challenge. For instance, the challenge in \bc is to come up with a \textit{nonce} that when hashed with block data produces a hash value that is less than a target threshold set by the system. All peers in the system use their computational power to solve the mathematical challenge. The peer who comes up with the solution wins the block race and mines a new block. Once a block is broadcast to the network, each peer verifies the solution and appends the block to his Blockchain. The probability of winning a block race is proportional to the computational power of participants. At the same time, there is a time restriction on the block mining~\cite{FullmerM18,BartolettiLP17}. In Bitcoin, the block time is set to 10 minutes. In other words, the network expects a new solution to the block puzzle after every 10 minutes. However, as the computational power increases, the chance of discovering a new block under 10 minutes increases. To address that, the network dynamically adjusts the difficulty of the challenge according to the change in the computational power of the miners. Oftentimes, more than one miner can come up with a valid solution leading to Blockchain forks and stale and orphaned blocks, which we discuss in \autoref{sec:bcstructure}.

In~\autoref{fig:overview}, we illustrate the transaction life-cycle in a proof-of-work (PoW)-based Blockchain application. User A (sender) generates a transaction for user B (receiver). The transaction is broadcast to the entire peer-to-peer network where it is temporarily stored in a transaction repository known as the memory pool (mempool). In a peer-to-peer network, the mempool is a space allocated in the RAM of a full node that stores and relays transactions to other peers. To maintain the state of the Blockchain, there are special nodes in the network known as the miners or verifiers, responsible for verifying transactions and computing a block. The miners query the mempool and select the transactions of their choice to put into blocks. Usually transactions pay a mining fee which can be viewed as an incentive given to the miners to mine the transaction. Naturally, miners give priority to the transactions that pay higher mining fee. Transactions that are not selected by the miners, stay in the mempool until some other miner selects them for a new block. Transactions that do not get mined for a long time, eventually get discarded.

\begin{table*}[]
\centering
\caption{An overview of popular consensus algorithms used in Blockchains. Notice that public and permissioned Blockchains using PoW, PoS, and DPoS have high scalability, low throughput, and high confirmation times. In contrast, permissioned Blockchains using PBFT and RAFT have low scalability, low confirmation time, and high throughput.  }
\label{tab:ca}
\begin{tabular}{l|c|c|c|c|c}

\textbf{Properties}                                                          & \textbf{PoW}       & \textbf{PoS}    & \textbf{DPoS}   & \textbf{PBFT}     & \textbf{RAFT}     \\ \hline

\textbf{\begin{tabular}[c]{@{}l@{}}Blockchain Type\\ \end{tabular}}         & Permisssionless & Permissionless & Permissionless & Permissioned    & Permissioned    \\ \hline

\textbf{\begin{tabular}[c]{@{}l@{}}Participation Cost\\ \end{tabular}} & Yes & Yes & Yes & No    & No    \\ \hline

\textbf{\begin{tabular}[c]{@{}l@{}}Trust Model\\ \end{tabular}} & Untrusted & Untrusted & Untrusted & Semi-trusted    & Semi-trusted    \\ \hline

\textbf{Scalability}                                                         & High               & High            & High            & Low               & Low               \\ \hline
\textbf{Throughput}                                                          & \textless{}10      & \textless{}1,000 & \textless{}1,000 & \textless{}10,000 & \textgreater{}10,000 \\ \hline
\textbf{\begin{tabular}[c]{@{}l@{}}Byzantine Fault Tolerance\\ \end{tabular}} & 50\%               & 50\%            & 50\%            & 33\%              & --                \\ \hline
\textbf{\begin{tabular}[c]{@{}l@{}}Crash Fault Tolerance\\ \end{tabular}}     & 50\%               & 50\%            & 50\%            & 33\%              & 50\%              \\ \hline
\textbf{\begin{tabular}[c]{@{}l@{}}Confirmation Time\\\end{tabular}}         & \textgreater{}100s & \textless{}100s & \textless{}100s & \textless{}10s    & \textless{}10s    \\ \hline

\end{tabular}
\end{table*}

\BfPara{Proof-of-Stake}The second most popular consensus algorithm in public Blockchains is proof-of-stake (PoS)~\cite{ThinDBD18,GuiHT18}. PoS was introduced to address the energy inefficiency of PoW. In PoS, the mining power of a user is determined by the total number of coins he owns. For each new block, an auction is carried out to select the candidate miner. Users place a bid on the block and the one with the highest bid is selected as a miner. Therefore, in contrast to PoW, the hashing power is replaced by the volume of assets owned by the user. The more coins a users owns, the higher his chances of winning the block race. The replace of energy intensive mining with stake-based mining, makes PoS energy efficient and secure against the majority attacks (\autoref{sec:51}). Unlike PoW, in PoS, all the cryptocurrency tokens are released prior to creation of the genesis block~\cite{DuongCFZ18}. Therefore, when a new block is mined, it does not introduce new coins in the system. However, miners are rewarded with transaction fee for their contributions.

\begin{figure}[t]
\begin{center}
\includegraphics[ width=0.45\textwidth]{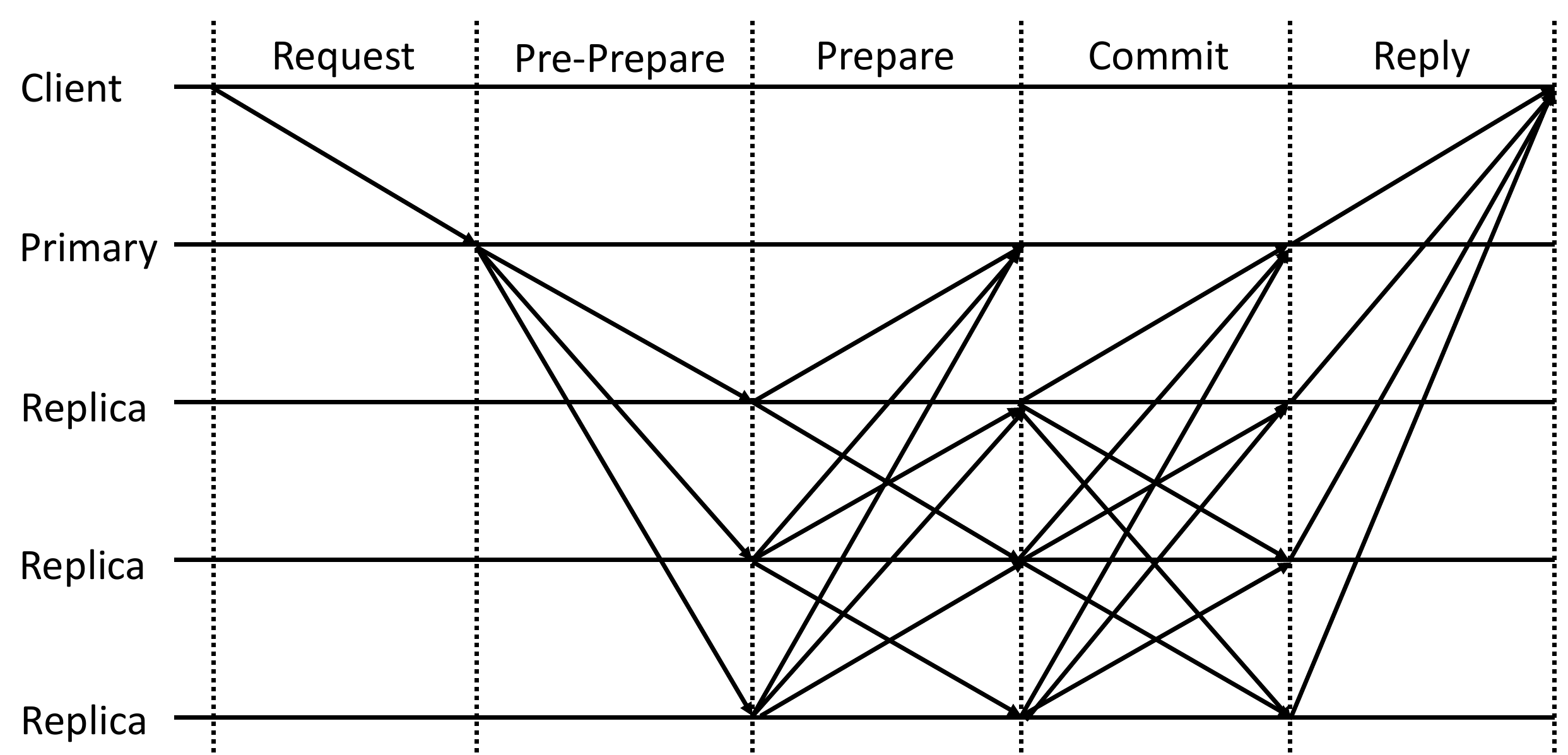}
\caption{Overview of PBFT protocol. The client issues a transaction to the primary replica. The primary replica then forwards it to other replicas who jointly execute a four-phased protocol and approve the transaction. Assuming  $f$ faulty replicas, the primary would require confirmation from at least $3f+1$ replicas. PBFT is employed in permissioned and private blockchains.   }
\label{fig:pbft}
\end{center}
\end{figure}

\BfPara{PBFT} The third most popular Blockchain consensus protocol is called the practical byzantine fault tolerance (PBFT)~\cite{AngelisABLMS18,SukhwaniMCTR17} protocol. PBFT is widely used in private and permissioned Blockchains, where the network has a stronger trust model compared to PoS and PoW. In PBFT Blockchains, the system is transposed into a group of active and passive replicas. Among the active replicas, a primary replica is selected who receives transactions from a client and sends them to the active replicas for execution. The process of execution is carried out in four stages, namely, pre-prepare, prepare, commit, and reply stage. In the pre-prepare phase, primary sends transactions to all the active replicas. In the prepare and commit phase, each active replica signs the transaction and exchange it with all the other replicas. In the reply stage, all the active replicas send their response to the primary replica. The primary collects all the signed transactions and puts them in a block. In \autoref{fig:pbft}, we show the transaction verification process in a PBFT Blockchain. Notice that compared to PoW and PoS, PBFT has a higher message complexity.

\begin{figure}[t]
\begin{center}
\includegraphics[ width=0.45\textwidth]{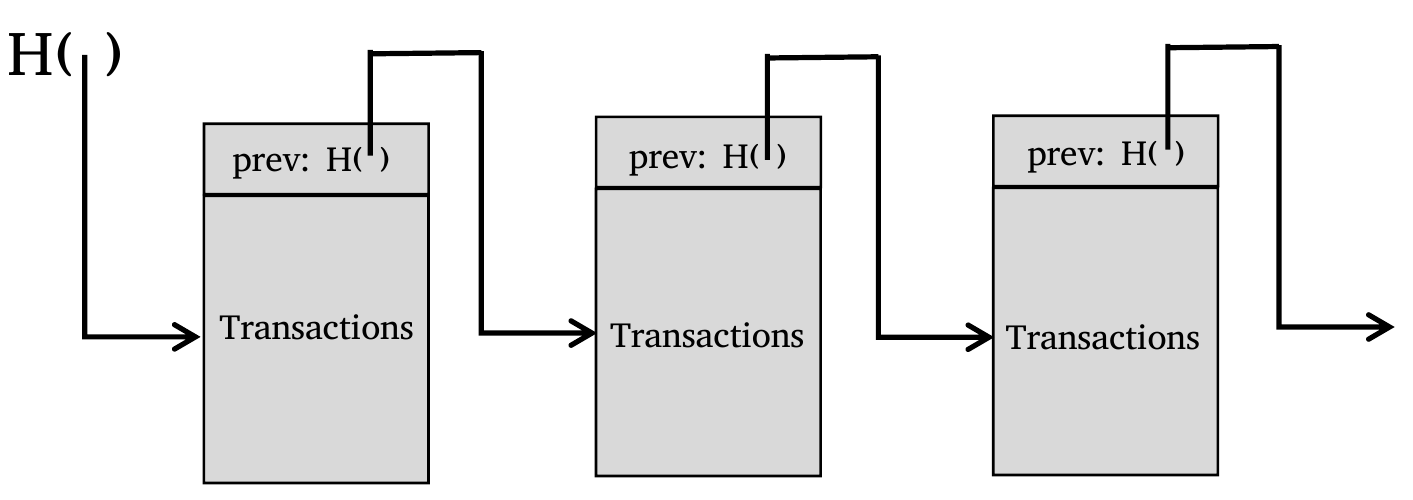}
\caption{Cryptographic constructs of blocks in a Blockchain. Notice that the entire hash of the previous block goes into the header of the next block. Therefore, if an attacker makes changes in the data of a block, he will be required to change data in all subsequent blocks and correctly execute the consensus protocol for each block. Since this is infeasible in practice, therefore Blockchains are considered tamper-proof.   }
\label{fig:hashing}
\end{center}
\end{figure}

In~\autoref{tab:ca}, we compare the popular consensus algorithms used in the Blockchain applications. Notice that permissionless Blockchains have low throughput and high confirmation time. Bitcoin has transaction throughput of 3--7 transactions per second. In contrast, permissioned Blockchains have a high throughput and and low confirmation time. In terms of security, PBFT has low fault tolerance ($\approx$ 33\%) compared to PoW and PoS ($\approx$ 50\%). However, since permissioned Blockchains have a stronger trust model, therefore, they are less vulnerable to adversarial attacks. It can also be observed in~\autoref{tab:ca}, public Blockchains are more scalable than private Blockchains~\cite{KimKC18,ChauhanMVM18}. This can be attributed to the message complexity involved in the transaction verification and the tolerance for Byzantine nodes. Since PBFT has high message complexity and low Byzantine fault tolerance, therefore, it cannot scale well beyond few hundred nodes. Therefore, each consensus scheme has its own benefits and limitations. Therefore, depending upon the application model, a consensus scheme can be selected to meet the requirements.

\subsection{Blockchain Structure} \label{sec:bchst}
While the consensus schemes in Blockchains may vary, the cryptographic constructs of Blockchain are fundamentally the same across all applications~\cite{GarayK18,CachinV17}. Each block in a Blockchain consists of a header and a payload. The header includes the primary information, such as the hash of the previous block, the merkle root, and the block timestamp. The hash pointer connects each block to the previous block, thereby forming a chain. Since hash functions are one-way and are collision resistant, Blockchain benefits from their properties to become immutable and tamper-proof~\cite{KonashevychP18,ChenLLLD18}. In \autoref{fig:hashing}, we illustrate this model of Blockchain, where blocks are linked through hash functions. 

In a Blockchain application, all nodes are connected in a peer-to-peer architecture. This means that they use the gossip protocol to communicate information, including transactions and blocks. Ideally, each peer is expected to maintain a copy of Blockchain. However, due to the append-only model, the growing size of Blockchain can put space constraint at the node. To address that, various Blockchain applications allow the segmentation of nodes into full nodes and lightweight nodes. The full nodes maintain a complete copy of Blockchain and participate in transaction and block propagation. On the other hand, the lightweight nodes only keep the block header for the verification of a newly published block.

As stated earlier, several attacks on Blockchain technology are related to the constructs of the Blockchain itself, the behavior of certain miners, and the peer-to-peer architecture it is built upon. In the subsequent sections, we explore the possible attacks associated with the Blockchain structure, attacks associated with the peer-to-peer architecture used in the \bcc system, and attacks associated with the application services that use \bcc technology (\ie {} \bc or Ethereum). We also supplement each section with possible countermeasures that have been proposed by researchers to address those attacks.

\section{Blockchain Structure Attacks} \label{sec:bcstructure}

In this section, we look at the attacks related to the design constructs of the Blockchain. These attacks emerge from the potential vulnerabilities of the Blockchain structures and as such, they can compromise any Blockchain-based application.

\begin{figure}[t]
\begin{minipage}{0.48\textwidth}
\begin{center}
\tikzstyle{int}=[draw, fill=blue!20, minimum size=2em]
\tikzstyle{init} = [pin edge={to-,thin,black}]
{\footnotesize
\begin{tikzpicture}[node distance=1.2cm,auto,>=latex']
	\node [int,align=center] (a1) {Old\\Rules};
	\node [int,align=center] (a2) [right of=a1] {Old\\Rules};
	\node [int,align=center] (a3) [right of=a2] {Old\\Rules};
	\node [int,align=center] (a4) [right of=a3] {Old\\Rules};
	
	\node [align=center] (a) [left of=a1, node distance=1.0cm] {Old\\Version};
	    
	\node [int,align=center] (b1) [below of=a3] {New\\Rules};
	\node [int,align=center] (b2) [right of=b1] {New\\Rules};
	\node [int,align=center] (b3) [right of=b2] {New\\Rules};
	\node [int,align=center] (b4) [right of=b3] {New\\Rules};
	
	\node [align=center] (b) [left of=b1, node distance=1.2cm] {New\\Version};

	\path[->] (a1) edge (a2);
	\path[->] (a2) edge (a3);
	\path[->] (a3) edge (a4);
	
	\path[->] (a2) edge node [midway, above, sloped] {Fork} (b1);
	
	\path[->] (b1) edge (b2);
	\path[->] (b2) edge (b3);
	\path[->] (b3) edge (b4);
	
\end{tikzpicture}
\caption{Hard Fork resulting from set of peers following conflicting rules due to different client software versions. Hard forks can be irreversible at times and may lead to a permanent split in the Blockchain application. }
\label{fig:bf}
}
\end{center}
\end{minipage}
\end{figure}
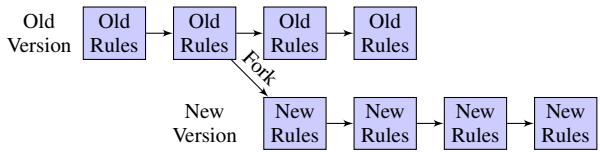

\subsection{Blockchain Forks} \label{sec:fork} A fork represents a condition in which nodes in the network have diverging views about that state of the Blockchain persisting over long periods of time or even indefinitely. These forks can be created unintentionally through protocol malfunctions or incompatibilities in client software upgrades. Forks can also be caused by malicious intents such as implanting ``Sybil nodes'' that follow conflicting validation rules or by carrying out ``selfish mining'' in race conditions as discussed further in~\autoref{sec:selfish}. Another form of fork occurs when users of a Blockchain application create a child application from the parent application. For example, in 2017, a group of Bitcoin developers decided to increase the block size limit from 1MB to 8MB by developing a new Bitcoin client that was capable of accepting 8MB blocks. However, their proposal was not accepted by the majority of users, therefore, they created a hard fork on Bitcoin and released a new cryptocurrency called Bitcoin Cash. Bitcoin Cash was the child application of the parent Bitcoin, with new rules and regulations. Therefore, forks can also be created to launch a new application.

\vspace{6mm}
\begin{vtimeline}[description={text width=5.5cm}, 
 row sep=2.2ex, use timeline header, timeline title={Major Bitcoin Forks}]
Jan 3, 2009 & Bitcoin genesis block established\endlr
Dec 27, 2014 & Bitcoin XT forked on Bitcoin Core\endlr
Jan 15, 2016 & Bitcoin Unlimited launched\endlr
Feb 10, 2016 & Bitcoin Classic forked on Bitcoin Core\endlr
Aug 1, 2017 & Bitcoin Cash launched\endlr
Aug 23, 2017 & Segregated Witness (Segwit) fork\endlr
Nov 1, 2017 & Bitcoin Gold launched\endlr
Nov 15, 2017 & Segwit2x fork\endlr
Nov 28, 2017 & Protest fork\endlr
\end{vtimeline}

Intentional forks can either be soft or hard, the latter of which occurs when new blocks that the network accepts appear invalid to pre-fork nodes.  Soft forks, however, occur when some blocks appear invalid to post-fork nodes. In either case, a Blockchain fork represents an inconsistent state that can be exploited by adversaries to cause confusion, fraudulent transactions, and distrust within network \cite{kwon2017selfish}. 

Figure \ref{fig:bf} demonstrates a hard fork example that results from peers following conflicting rules about the state of Blockchain. Such hard forks may lead to a split in cryptocurrency. A major hard fork on Bitcoin occurred during August 2017, which led to the creation of Bitcoin Cash \cite{bitcoincash}. Another hard fork on Bitcoin occurred during October 2017, when Bitcoin Gold \cite{goldbitcoin} was created. Some other notable forks in Bitcoin include Bitcoin Classic, Bitcoin XT, and Bitcoin Unlimited. However, due to insufficient user-base and miners, they could not succeed as a separate cryptocurrency.

When hackers stole more than one third of the total digital cash owned by ``The DAO'' \cite{DAO}, Ethereum used a hard fork to roll back transactions and retrieve millions of dollars' worth of ether (the ``fuel'' for the Ethereum network). However, this required consensus by the majority of nodes in the network. In such a scenario, if a consensus delay happens due to a majority attack or a DDoS event, fraudulent activities become somewhat difficult to deal with and prolonged delays can ultimately cause devaluation of \cc. In November 2017, the second version of Segregated Witness (SegWit2x) hard fork was proposed in Bitcoin, which aimed to increase the block size to 2MB. However, due to lack of consensus by the majority, the planned hard fork was canceled. In Timeline 1, we provide a list of major forks on Bitcoin. These forks resulted from a group of miners introducing new rules and a faction of peers switching to those rules. All these forks introduced a new version of Bitcoin. This is, we note that a fork may diminish if peers discontinue to follow the new rules and switch back to the old ones. For instance, this has been witnessed in SegWit fork. Initially, a faction of network peers switched to SegWit version of Bitcoin, however, when they moved back to the old version in a protest, the fork ended.

\subsection{Stale Blocks and Orphaned Blocks} \label{sec:stale}
Two forms of inconsistencies can occur with the consensus process that can leave valid blocks out of the Blockchain. The first form is a ``stale block'', which is a block that was successfully mined but is not accepted in the current best Blockchain (\ie the most-difficult-to-recreate chain). Stale blocks occur mostly in the public Blockchains due to race conditions. In race conditions, the miners actively try to find the next block, and it is possible that two or more miners can come up with a valid solution. The network eventually accepts one of the winning blocks and discards the rest. As a result, the all other valid blocks unaccepted become stale blocks as they do not get attached to the main Blockchain. We will see in~\autoref{sec:selfish} that a form of Blockchain attack known as ``selfish mining'' can also lead to the creation of stale blocks in the network, which deprives an honest miner of its reward.

The other form of inconsistency is an ``orphaned block'': a block whose parent block's hash field points to an unauthentic block that is detached from the Blockchain~\cite{de2012bitcoin}. These inconsistencies can be introduced by an attacker or caused by race conditions in the work of the miners. Stale blocks may be initially accepted by the majority of the network, but they can be rejected later when proof of a longer Blockchain (\ie the current best) is received that does not include that block. 

\begin{figure}[t]
\begin{minipage}{0.48\textwidth}
\begin{center}
{\small
\tikzstyle{int}=[draw, align=left, fill=blue!20, minimum size=2em]
\tikzstyle{init} = [pin edge={to---,thick,black}]
\tikzstyle{intx}=[draw, fill=blue!20, minimum size=2em, dotted]

\begin{tikzpicture}[node distance=1.8cm,auto,>=latex']
    \node [int,align=center] (a)  {Block 1  }; 
    \node [int,align=center] (b) [right of=a] {Block 2 } ; 
     \node [int,align=center] (c) [right of=b]  {Block 3  }; 
     
     \node [int,align=center] (d) [above right of=b, fill=red!20]  {Block 5   } ; 
     
     \node [int,align=center] (e) [below right of =a,fill=red!20 ]  {Block 2 } ; 
    
   \node [intx,align=center] (f) [above right of =a,fill=red!20 ]  {Block 4 } ;

    \path[->] (a) edge (b);

    \path[->] (a) edge (e);
    
    \path[->] (b) edge (c);
    
      \path[->] (f) edge (d);
    \draw[densely dotted] (c) edge (f);

\end{tikzpicture}
\caption{Stale vs. orphaned blocks. Note that the stale block (block 2, bottom, and block 4) are valid but they are not part of the Blockchain. Orphaned block (block 5) does not have its parent block (block 4) in the Blockchain.}
\label{fig:ob}}
\end{center}

\end{minipage}

\end{figure}
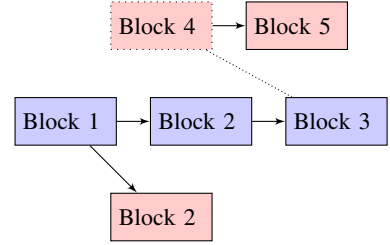

\begin{figure}[t]
\begin{center}
\includegraphics[ width=0.48\textwidth]{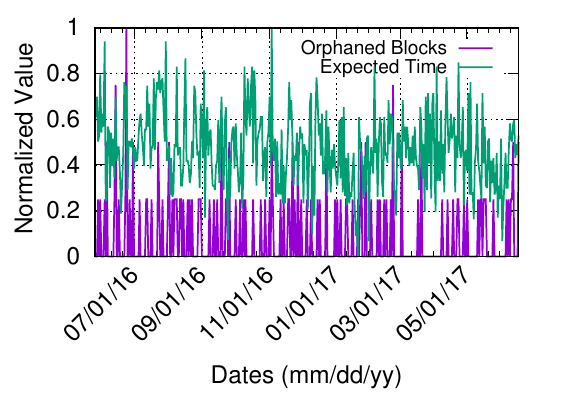}
\caption{Orphaned Blocks in Bitcoin and Uncle blocks in Ethereum over the last two years. Notice that in Bitcoin, the rate of orphaned blocks has reduced.   }
\label{fig:orphans}
\end{center}
\end{figure}

Figure \ref{fig:ob} demonstrates a chain where stale and orphaned blocks can be found. The first orphaned block in \bc was found on March 18, 2015, and that was the beginning of a period in which most orphaned blocks were created. The trend reduced in 2016, and from June 2017 to the date of this paper, no orphaned block has been added to the list \cite{block-explorer}. Orphaned blocks are more frequently found in cryptocurrencies where average block computation time is small. In \autoref{fig:orphans}, we plot the number of orphaned blocks that occurred in Bitcoin and Ethereum from July 2016 to May 2018. In Ethereum, the orphaned blocks are called Uncle blocks. The data in the figure has been normalized using min-max normalization  to scale the data in the range $[0,1]$. The min-max scaling is conducted as $z  = \frac{x_i - \min (x)}  {\max (x) - \min (x)}$. It can be observed from the figure that as of June 2017, no orphaned block has been found in Bitcoin. On the other hand, in Ethereum, Uncle blocks have increased since November 2017.

In cryptocurrencies such as Ethereum and Bitcoin, the difficulty is a measure of how long it takes to compute a block, which is defined by a target value set by the network~\cite{difficulty}. Based on the hashing power, the {\em target} is adjusted to keep block time under a predefined range (10 minutes for Bitcoin and 12 seconds for Ethereum). The difficulty is recomputed based on the hashing power and the time taken by a series of previous blocks: if hashing power increases, the probability of finding a block under the expected time increases. 

To adjust the probability, the difficulty is raised by increasing the target value. In~\autoref{equ:prob}, we show how the expected time to compute a block $E(T)$ varies with the difficulty $D$ and hash rate of the network $H_r$. Here, $E(T)$ is measured in seconds, $D$ is the number of hashes required to solve the current target, and $H_r$ is measured in hashes/second, that a target device can produce over a given string. $H_r$ is the aggregate hashing power of all the miners $H_i$ for $i = 1,2,\dots,n$.  In~\autoref{equ:time}, we calculate the time $T_b$ (seconds), it takes for a single miner in $H_i$ to compute a block, given a fixed block time set by the network $T_n$. For Bitcoin and Ethereum, the average block computation time $T_n$ is 600 seconds and 12 seconds, respectively. 

\begin{align} 
  \label{equ:prob}  &H_r  = \sum_{i=1}^{n} H_i,  &E(T) =  \frac{D}{H_r}\\  
  \label{equ:time} &T_b = \frac{T_n \times H_r}{H_i} &T_b = \frac{T_n \times \sum_{i=1}^{n} H_i}{H_i}
\end{align}

\begin{figure*}
\hfill

\begin{subfigure}[Change in difficulty and hash rate of Bitcoin network during 2016-17 \label{fig:bfive}]{\includegraphics[width=5.5cm]{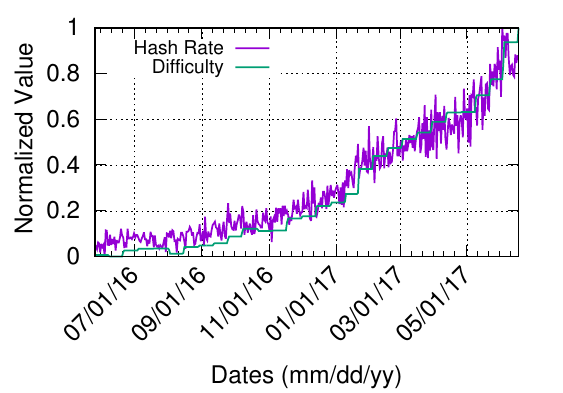}} 
\hfill
\end{subfigure}
\begin{subfigure}[Expected time $E(T)$ calculated from~\autoref{equ:time} plotted against the actual time  \label{fig:bsix}]{\includegraphics[width=5.5cm]{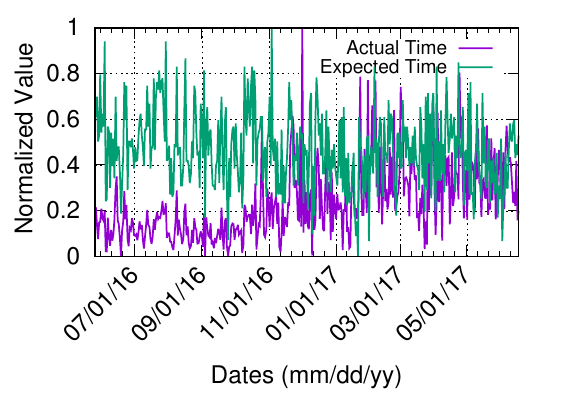}}
\hfill
\end{subfigure}
\begin{subfigure}[Orphaned Blocks per day plotted against the expected block time. \label{fig:bseven}]{\includegraphics[width=5.5cm]{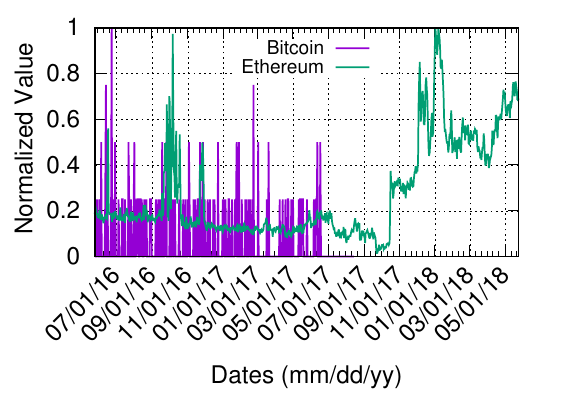}} 
\hfill
\end{subfigure}

\begin{subfigure}[Change in difficulty and hash rate of Ethereum network during 2015-18. \label{fig:five}]{\includegraphics[width=5.5cm]{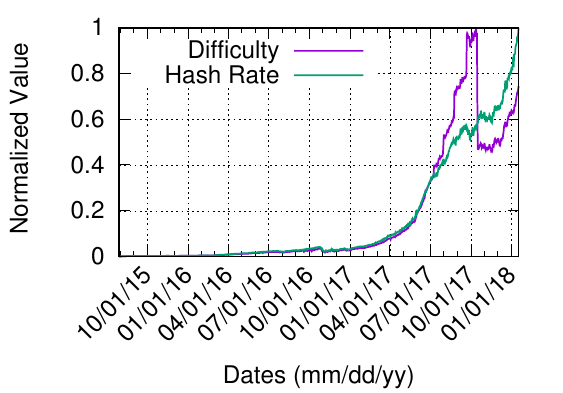}} 
\hfill
\end{subfigure}
\begin{subfigure}[Expected time $E(T)$ calculated from~\autoref{equ:time} plotted against the actual time  \label{fig:six}]{\includegraphics[width=5.5cm]{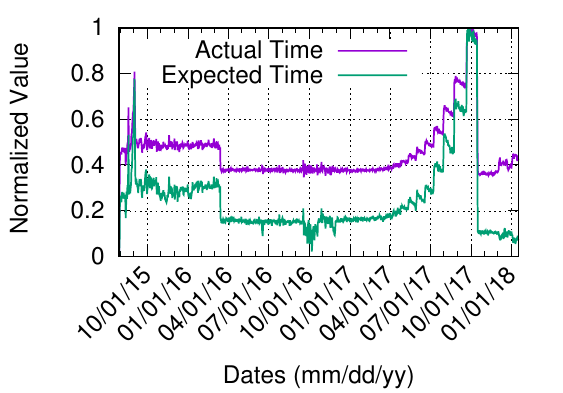}}
\hfill
\end{subfigure}
\begin{subfigure}[Uncle Blocks per day plotted against the expected block time. \label{fig:seven}]{\includegraphics[width=5.5cm]{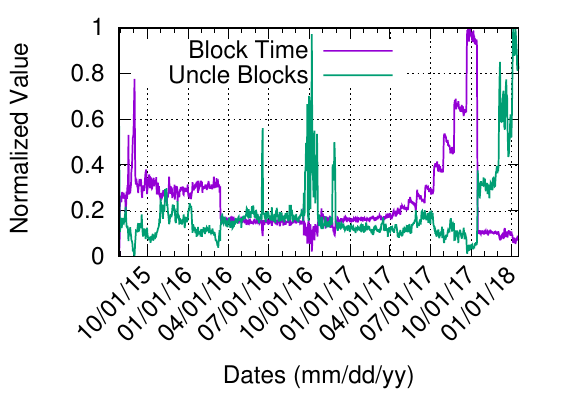}} 
\hfill
\end{subfigure}

\caption{Effect of hash rate and difficulty on the rate of orphaned blocks in Bitcoin and uncle blocks in Ethereum. For Ethereum, notice that when the difficulty sharply decreases with constant hash rate around October 2017, the expected and the actual time of block computation decreases sharply. As a result, the number of Uncle blocks increases. The sharp decrease in the difficulty is associated to a byzantium fork that reduced block rewards per block.} 
\label{fig:eight}
\end{figure*}

From~\autoref{equ:prob}, it can be observed that when $H_r$ remains constant and the difficulty $D$ is reduced, the expected block time $E(T)$ decreases. Intuitively, lower $E(T)$ means that in a defined network time $T_n$, more blocks will be produced. However, in the Blockchain, only one block can be accepted. Such a situation will lead to more orphaned blocks in the system. In~\autoref{fig:eight}, we plot difficulty, hash rate, block time and orphaned blocks (also called uncle blocks) in Ethereum. It can be noted in~\autoref{fig:seven} that as the expected block time (from~\autoref{equ:time}) decreases, the number of orphaned and uncle blocks increases. In Ethereum, this trend is high due to short block intervals which increase the possibility of block collision. Orphaned blocks may also occur due to unpredictable delays in block propagation. A valid block may not reach majority of the network peers due to network churns and propagation delays. In contrast, a competing block is able to easily propagate through the network and get accepted by the majority. Therefore, network behavior and delay distribution may also affect the number of orphaned blocks in a Blockchain system \cite{de2012bitcoin}.

\subsection{Vulnerabilities in Consensus Mechanism} \label{sec:cons}
\subsubsection{Proof-of-Work} The most widely used consensus protocol in cryptocurrencies is proof-of-work (PoW) which serves as an evidence for the effort put behind the computation of a valid block. As outlined in \autoref{equ:prob}, the effort for computation of a block can be characterized as the number of hashes required to meet the difficulty parameter $D$ set by the network. As the aggregate hash power of the network $H_r$ increases, the difficulty is raised to keep the standard block time $T_n$ within a defined range (10 minutes for Bitcoin). 

\begin{table}[t]
\centering
\caption{Evolution of mining hardware. Since 2014, only ASIC chips, with upgraded versions, are being used for mining.}
\label{tab:bc}
\scalebox{1}{
\begin{tabular}{|l|c|c|l|} 
\hline
\textbf{Type} & \textbf{Model} & \textbf{\begin{tabular}[c]{@{}l@{}}Hash Rate\\ (MH/s)\end{tabular}} & \textbf{Year} \\ \hline
CPU           & Xeon E5530     & 7.14                                                               & 2009          \\ \hline
GPU           & Radeon 5890    & 245                                                                & 2010          \\ \hline
GPU           & Radeon 6990    & 800                                                                & 2011          \\ \hline
FPGA          & Xilinx Spartan & 245                                                            & 2012          \\ \hline
FPGA          & Xilinx Spartan & 850                                                                & 2012          \\ \hline
ASIC          & ASIC 130nm     & 12K                                                                & 2013          \\ \hline
ASIC          & ASIC 28nm      & 500k                                                               & 2014          \\ \hline
ASIC          & ASIC 20nm      & 750k                                                               & 2014          \\ \hline
\end{tabular}}
\end{table}

In \autoref{fig:bfive} and \autoref{fig:five}, we show the increase in difficulty and the aggregate hash rate of Bitcoin and Ethereum, respectively. Since mining in PoW is a lottery-based system, miners use sophisticated hardware with high hash rate to increase their chances of winning the lottery. Among all PoW-based cryptocurrencies, Bitcoin has the maximum hash rate. In particular, and since 2010, miners in Bitcoin have switched from Central Processing Unit (CPU), to graphics processing unit (GPUs) in 2011, to Field Programmable Gate Array (FPGA) in 2012--13, and finally to Application Specific Integrated Circuit (ASIC) chips since 2014 to date~\cite{greene_2018}. We show this evolution of Bitcoin hardware, along with the hash rate, in \autoref{tab:bc}.

\begin{figure}[t]
	\centering
	\includegraphics[width=0.9\linewidth]{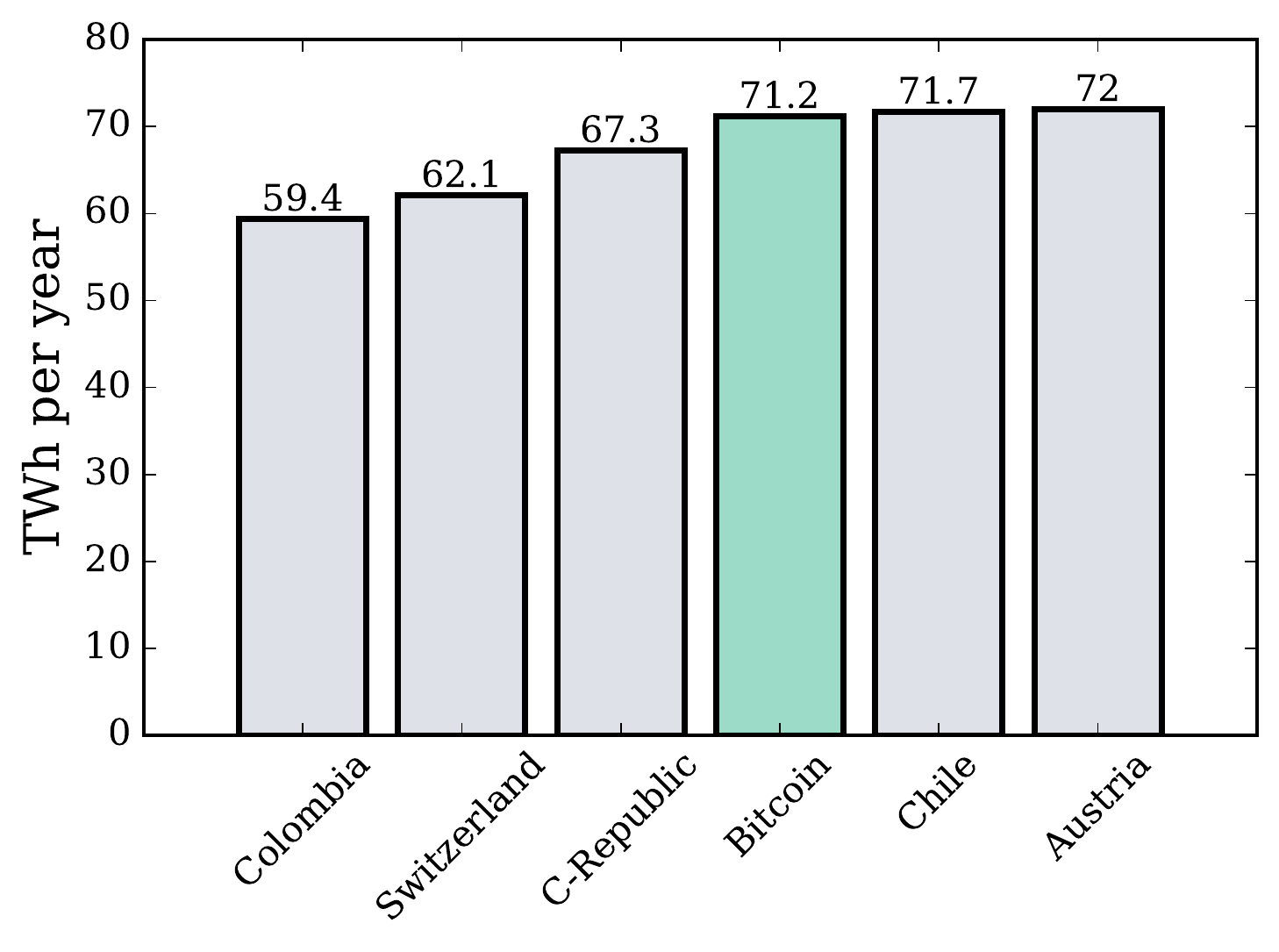}
	\caption{Energy Profile of Bitcoin and other countries. C-Republic refers to Czech Republic. Note that Bitcoin's consumption is compare able to countries.}
	\label{fig:ec}
\end{figure}

One of the major problems with PoW is the excessive waste of energy to find a valid solution \cite{BeckerBHHRB13}. At present, Bitcoin and Ethereum use 71.12 Terawatt-hours and 4.2 Terawatt-hours (TWh) of electricity per-year, respectively, to find hashes required for valid PoW \cite{de2018bitcoin,kang2016bitcoin,digiconomist}. In \autoref{fig:ec}, we show the electricity consumption of Bitcoin compared to several countries. Other than the excessive consumption of electricity, centralization of hashing rate among a few mining pools makes the Blockchain application vulnerable to attacks including the majority attacks and double-spending (discussed in \autoref{sec:51} and \autoref{sec:double-spending}), whereby if a miner acquires the majority of a network's hash rate, the miner will be able to gain  control over the system.  

\subsubsection{PoS}
(PoS) was introduced by King and Nadal in 2012 \cite{king2012ppcoin} to make Blockchain applications more energy-efficient and raise the cost of a majority attack. Unlike PoW, which is lottery-based, PoS uses a stake-based deterministic approach to select a validator and to publish a new block \cite{GaziKR18}. In this approach, the validator is chosen by a bidding process, whereby candidate validators make a bid of their stake. The stake is the balance owned by the candidate validator and is used to deter cheating in the system. The candidate with the highest bid is chosen to mine the next block and if he tries to trick the system with bogus transactions he risks losing his committed stake (balance). The process is deterministic since a validator is chosen prior to each bidding process. Therefore, blocks are published on their expected time without time deviations or delays. Moreover, to launch a majority attack on a PoS-based \cc, the attacker is required to acquire more than 50\% of the \cc tokens \cite{KiayiasKRDO16}. While it is relatively easier to acquire 50\% hash rate in PoW, it is difficult to obtain 50\% coin. Therefore, compared to PoW, the cost for launching a majority attack in PoS application is relatively high, which makes the attack less feasible. 

Although PoS serves as a ``green'' mining alternative of PoW and raises the attack cost for the majority attacks, it has some major caveats that have prevented its widespread adoption by the Blockchain community. In PoS, a rich validator may keep on winning the bid for the next block to be validated, and accumulate the block reward. As such, the rich validators in the system gets richer for block confirmation, which makes PoS applications centralized around those validators. This challenges the fundamental premise of Blockchain technology as a decentralized system~\cite{chang2017blockchain}. Moreover, unlike PoW, in which miners with limited resources may still have a chance of winning the lottery, small bidders in PoS are certain to lose the bid for each coming block.

\subsubsection{PBFT}
As pointed out in \autoref{sec:bchst}, in PBFT-based private Blockchains, the system is grouped into a set of replicas that process transactions and contribute towards the block formation~\cite{BanoSBAMMD17,Yang18}. The primary replica is responsible for ordering transactions and obtaining approvals from other replicas. Once sufficient approvals are received, the primary computes a block and broadcasts it to the network. PBFT is considered to be energy efficient with high transaction throughput. However, it works under the assumption that the primary replica faithfully executes the protocol and does not tamper with the ordering of transactions and blocks. This assumption may lead to a vulnerabilities in the permissioned Blockchains. If the primary replica is compromised it may:
\begin{enumerate}
    \item discard the approvals obtained from other replicas and prematurely abort the execution, 
    \item rearrange the sequence of transactions to delay the verification process and block generation, 
    \item withhold transactions or blocks from other replicas, and/or
    \item invalidate transactions even after obtaining approvals.
\end{enumerate}
As such, private Blockchains always are subjected to the risk of a malicious primary who may compromise the system. However, since the identity of the primary is usually known to everyone, malicious activities of a primary can be tracked back, eventually.

Another key challenge in PBFT-based private Blockchains is their limited scalability and low tolerance to Byzantine nodes. Low scalability results from the $O(n^{2})$ message complexity associated with the processing of a single transaction, as shown in \autoref{fig:pbft}. In PBFT, transaction execution is carried out in four phases, namely pre-prepare, prepare, commit, and reply. In the prepare and commit phase, each peer is required to send message to every other peer in the network. In aggregate, this leads to enormous communication overhead which cannot be expected to work efficiently at a large scale. Therefore, when the network size grows, the performance of PBFT is degraded significantly. This is the key reason why PBFT-based private Blockchains suffer from low scalability.

Finally, another limitation of PBFT-based private Blockchains is their low fault tolerance. Each transaction requires approval from $3f+1$ replicas, where $f$ is the number of faulty replicas or Byzantine nodes. In comparison with PoW and PoS, where the network can withstand up to 50\% malicious entities, PBFT can only tolerate 33\% malicious replicas. Provided that PBFT already suffers from low scalability, a lower fault tolerance increases the opportunity for an adversary to place malicious replicas in the network. Currently, Bitcoin has over 10,000 active full nodes~\cite{bitnodes_18}. This means that it can tolerate up to 5,000 faulty nodes. The cost of compromising 5,000 nodes is high, and the attack is therefore infeasible. However, in a PBFT-based private Blockchain that consists of a 100 nodes, an attacker can succeed only by controlling 33 nodes. Low fault tolerance is a major challenge in PBFT-based Blockchain applications. 

Considering the features and shortcomings of existing consensus algorithms, there is a need for new consensus mechanisms that are secure, scalable, and energy efficient. Currently, this remains an active research area, with some notable recent progress made in this direction~\cite{PassS18,Rocket18,BergerR18}.

\subsection{Countering Blockchain Structure Attacks} \label{sec:countering}
Resolving soft forks in a Blockchain network is a relatively easy process. All peers in the network can come to a consensus about the true state of the Blockchain and resume activities from there. Resolving hard forks can be challenging because conflicting chains can be lengthy with transaction activities dating back to the time of the conflict. Although the stakes of rolling back from a hard fork are high, they can be resolved by the same principle of consensus as discussed earlier. As was the case with Ethereum, a hard fork was used to retrieve money for the investors after ``The DAO'' was attacked \cite{DAO}. Ultimately, the process of solving a fork depends upon the agreement of peers in the network and their stake in the fork.

In Ethereum, uncle blocks are also rewarded and made part of the Blockchain. Recently, the number of orphaned blocks in \bc has decreased due to the shift towards highly centralized mining networks and thus reducing the probability of orphaned blocks prevalent in decentralized mining networks. However centralized mining has other issues such as unfairness in the network and the 51\% attack. The other solution to avoid stale or orphaned blocks involves dynamic adjustment of network's difficulty \cite{VelnerYJL_17}. In Bitcoin, the difficulty is adjusted every two weeks (2016 blocks). In the meantime, if there is a sharp increase in the hash rate of the network or more miners join in, then the expected time of finding new block decreases~\autoref{equ:time}. As a result, there is a higher likelihood of producing stale blocks. Therefore, a dynamic difficulty adjustment helps in reducing the number of stale and orphaned blocks. While there are effective techniques to counter forks and orphaned blocks, the area of consensus remains open. Research efforts need to be dedicated to make PoW more energy efficient, and PoS, more decentralized. In PBFT-based private blockchains, the key issue is limited scalability due to high message complexity. Moreover, PBFT has low fault tolerance which makes it vulnerable to attacks. In \autoref{sec:cp2p}, we provide more details about making PBFT more scalable and secure.

\section{Blockchain's Peer-to-Peer System}\label{sec:p2p}
The underlying peer-to-peer architecture is the primary reason why certain guarantees are provided by a Blockchain, including security and accessibility. Counter intuitively, this peer-to-peer architecture that the Blockchain resides on actually contributes to several attacks including selfish mining, the 51\% attack, DNS attacks, distributed denial-of-service attacks, eclipse attacks, fork after withholding attacks, and consensus delay. In this section, we explore how these attacks can compromise the Blockchain applications.

\subsection{Selfish Mining}\label{sec:selfish}
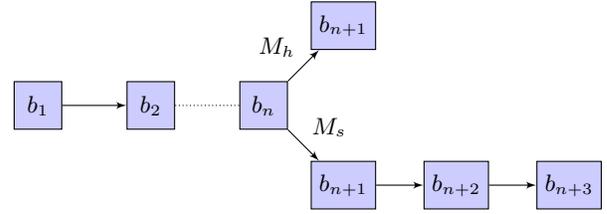
\begin{figure}[t]
\begin{minipage}{0.5\textwidth}
\begin{center}
{\small
\tikzstyle{int}=[draw, fill=blue!20, minimum size=2em]
\tikzstyle{init} = [pin edge={to-,thin,black}]
\begin{tikzpicture}[node distance=1.5cm,auto,>=latex']
    \node [int] (a) {${ b}_1$};
    \node [int] (b) [right of=a] {${ b}_2$};
     \node [int] (c) [right of=b] {${b}_n$};
     
     \node [int] (d) [above right of=c] {${b}_{n+1}$};
     
     \node [int] (e) [below right of=c] {${b}_{n+1}$};
          \node [int] (g) [right of=e] {${b}_{n+2}$};
                    \node [int] (h) [right of=g] {${b}_{n+3}$};
    \path[->] (a) edge (b);
    \path[->] (c) edge node {${M}_h$} (d);
    \path[->] (c) edge node {${M}_s$} (e);
    \path[->] (e) edge (g);
        \path[->] (g) edge (h);
    \draw[densely dotted] (b) edge (c);
\end{tikzpicture}
\caption{Illustration of Selfish Mining. Selfish behavior of $M_s$ forks the chain at $b_{n+1}$ and discards $M_h$'s block. $M_h$'s block becomes a stale block.  }
\label{fig:sm}
}
\end{center}
\end{minipage}
\end{figure}

The selfish mining attack~\cite{eyal2014majority} is a strategy opted by certain miners who attempt to increase their rewards by deliberately keeping their blocks private \cite{LeelavimolsilpTS18,GrunspanM18,NayakKMS16}. Rather than releasing them to the public upon discovery, these selfish miners continue to mine their own private blocks to obtain a longer chain than the public \bcc. These activities lead to a block race between the public chain of honest miners and the private chain of selfish miners. Once the public \bcc starts approaching the length of their private chain, selfish miners release their blocks to claim block rewards. Having exceptional mining power may further help selfish miners win the block race.  In~\autoref{fig:sm}, we demonstrate how a selfish mining attack is carried out.

Consider a Blockchain with blocks ($b_1$, $b_2,\ldots,b_n$). Suppose an honest miner $M_h$ has successfully mined the next block $b_{n+1}$ and he publishes it. All the peers in the network validate and accept his block. At the same time, a selfish miner $M_s$ also computes the block $b_{n+1}$. Instead of publishing his block, $M_s$ chooses to withhold it and successfully mines two more blocks $b_{n+2}$ and $b_{n+3}$. Despite $M_h$ 's block being added to the Blockchain, we show that $M_h$ can still be cheated while having a majority of network's confidence in his block. 
Let the hash value of $M_h$'s block $b_{n+1}$ be lower than both the target threshold and $M_s$'s block $b_{n+1}$. If only these two blocks were presented to the network, $M_h$'s block would be chosen (due to its greater computational complexity) over $M_s$'s block  and appended to the public \bcc. 

However, after some time, $M_s$ releases all of his blocks $b_{n+1}$ , $b_{n+2}$, and $b_{n+3}$ and forks the Blockchain at $b_{n+1}$ . Due to the design protocols of Blockchain, the network will invariably shift to the longer chain belonging to $M_s$ and discard the block $b_{n+1}$ of $M_h$. The effort put forth by $M_h$ in computing his block will be wasted due to selfish behavior of $M_s$. The incentive in adopting this selfish mining strategy is maximizing block rewards by publishing a longer chain. It should be noted that excluding the $M_h$'s block $b_{n+1}$ from the \bcc does not destroy the block, rather it leads to another significant problem in the network known as ``stale  blocks'' as shown in~\autoref{sec:stale}.

Selfish mining attacks can produce undesirable results for the rest of the network by invalidating the blocks of honest miners who contribute to the Blockchain. Furthermore, all the transactions in the honest miner's block also get rejected. In a situation where two selfish miners compete to add their chains to the network, the chances of a ``Blockchain fork'' arise~\autoref{sec:fork}. These forks can cause a delay of consensus in the network, which can further lead to other potential attacks such as ``double-spending'' and ``fork after withholding'', as discussed in~\autoref{sec:double-spending}. One selfish activity in the network has the potential to disrupt the overall network, and therefore it is imperative to study their relationship with one another.

\subsection{The Majority Attack}\label{sec:51} 

The majority attack also known as the 51\% attack is well known vulnerability in Blockchain-based applications that can be exploited when a single attacker, a group of Sybil nodes, or a mining pool in the network attains the majority of the network's hash rate to manipulate the Blockchain. With majority of network's hash rate, the attackers are able to 1) prevent transactions or blocks from being verified (thus making them invalid), 2) reverse transactions during the time they are in control to allow double-spending, 4) fork the main Blockchain and split the network, and 3) prevent other miners (verifiers) from finding any blocks for a short period of time. Under race conditions, the attackers with over 50\% hash rate are guaranteed to over take other miners and append their blocks in the Blockchain with high probability \cite{source}. Also, these blocks can possibly have fraudulent or double-spent transactions. For example, if an attacker performs a transaction in exchange for any product with Alice, it can replicate the same transaction with Bob and put it on the block. Transactions on Blockchains are not reversible, and only one transaction can be considered valid. In the following, we elaborate the prospects of double-spending with majority attacks along with the mathematical primitives.

\subsubsection{Caveats and realities} Mining pools do not always need 51\% of the network's hashing power to carry out the fraudulent activities. As such, even with less hashing power, similar objectives can be achieved with a significant probability of success. To understand this issue, consider the scenario in which a malicious mining pool with significant hash rate carries out a transaction $Tx$ with a receiver. At the same time, it generates a fraudulent double-spent transaction $Ty$ from the same parent transaction to trick the receiver. The receiver, on the other hand, waits for $k$ confirmations before releasing the product to the miner. The $k$ confirmations mean that $k$ subsequent blocks have been mined by the network after mining the transaction $Tx$. During this process, the malicious miner keeps mining blocks on his end with the double-spent transaction $Ty$ and hopes to fork the Blockchain after he receives the product from the recipient. By forking the chain, the malicious miner will be able to invalidate the chain with transaction $Tx$, and will replace it with his own chain with double-spent transaction $Ty$.

To launch a successful attack, the malicious miner needs to publish a longer chain with valid PoW so that the network switches to his forked version. Miner's success depends on his hash rate $x$ as a fraction of the network's hash rate and the number of confirmations $k$. To find the probability of success $P(s)$ for the attacker, let $x$ be the fraction of miner's hashing power and $y$ be the fraction of remaining hashing power, where $x+y = 1$ \cite{eyal2014majority}. The success probability is:
\[
P(s)= \begin{cases} 
{1} & \text{, if $x > y$}\\
(\frac{x}{y})^k &\text{, if $x < y$}\\
\end{cases}
\]

\begin{figure}[t]
\begin{center}
\includegraphics[ width=0.45\textwidth]{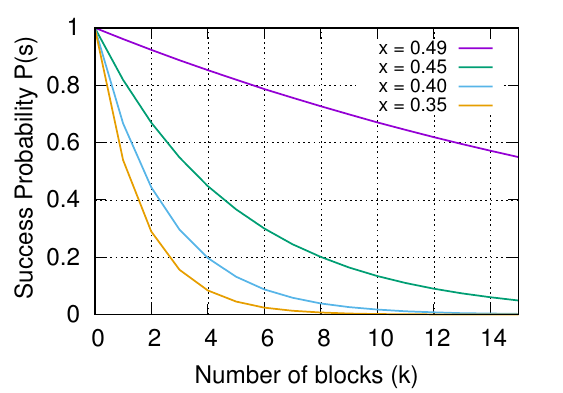}
\caption{Change in the success probability $P(s)$ of majority attack with varying hashing power $x$ and number of confirmations $k$. Notice that with 0 confirmations, an attacker can always double-spend with any magnitude of hash power. In that case we describe the user to be optimistic. }
\label{fig:51}
\end{center}
\end{figure}

\subsubsection{Numerical results} In \autoref{fig:51}, we show how $P(s)$ changes with varying hash rate. Note that if the miner acquires half of the network's hash rate, he can trick the recipient with 100\% success rate. Moreover, an attacker with hash rate less than 50\% can still succeed in forking the main chain and cheating the receiver. 

\subsubsection{Applications and implications} A Blockchain-based application for Internet of Things (IoT), known as ``The Tangle'' \cite{thetangle} can be theoretically compromised with one-third of the hash power. Bahack \etal \cite{bahack2013theoretical} show that the majority attacks are highly feasible with one quarter of the network's hashing power. There are online services such as Nicehash, that rent hashing power to miners on hourly basis \cite{nicehash_18}. 

\begin{table}[t]
\centering
\caption{Attack cost required to launch the 51\% attack on the top six Blockchain-based cryptocurrencies. Here Cap denotes the market cap in USD, Algo denotes the algorithm used for block consensus, and cost denotes the attack cost in USD for launching the 51\% attack for one hour. }
\label{tab:51}
\scalebox{0.95}{
\begin{tabular}{|l|l|l|l|l|}
\hline
{\sc System}         & {\sc CAP} & {\sc ALGO} & {\sc Hash Rate} & {\sc Cost} \\ \hline
{\sc Bitcoin}      & 112.7B            & SHA-256            & 35,604 PH/s        & 486K              \\ \hline
{\sc Ethereum}     & 49.5B             & Ethash             & 222 TH/s           & 347K              \\ \hline
{\sc B.Cash } & 14.9B             & SHA-256            & 5,023 PH/s         & 68K              \\ \hline
{\sc Litecoin}     & 5.7B              & Scrypt             & 327 TH/s           & 60K               \\ \hline
{\sc Dash}         & 2.1B              & X11                & 2 PH/s             & 15K               \\ \hline
{\sc Monero}       & 2.3B              & CryptoNight        & 365 MH/s           & 17K              \\ \hline
\end{tabular}}
\end{table}

A malicious mining pool can rent the computation power for a few hours and launch the majority attack on the targeted cryptocurrency. Since major blockchain systems have a high aggregate hash rate, the renting cost to launch the 51\% attack on them is (naturally) high. In \autoref{tab:51}, we outline the top six Blockchain-based cryptocurrencies, and the cost required to successfully launch the 51\% attack, based on data obtained from ``51crypto'' \cite{bc_community_18}. We notice that Dash with a market cap of 2.3 Billion USD can be compromised for one hour by spending only 17,000 USD ($8 \times 10^{-4}$\% of the market cap).

\begin{table*}[t]
\centering
\caption{Location of full nodes in three major cryptocurrencies. -- in Bitcoin refers to the nodes that use TOR services and their location cannot be identified. }
\label{tab:spa}
\scalebox{0.95}{
\begin{tabular}{|c|l|l|l|l|l|l|}

 \hline
    \multirow{2}{*}{} &
      \multicolumn{2}{c}{\textbf{Bitcoin}} &
      \multicolumn{2}{c}{\textbf{Ethereum}} &
      \multicolumn{2}{c|}{\textbf{Litecoin}} \\
 \hline
{Rank} & {Country} & {Nodes} & {Country} & {Nodes} & {Country} & {Nodes} \\ \hline
{1}    & Unite States     & 2445 (24.98\%) & United States    & 6549 (37.99\%) & United States    & 79 (24.38\%) \\ \hline
{2}    & Germany          & 2445 (24.98\%) & China            & 2202 (12.77\%) & Russia            & 36 (11.12\%) \\ \hline
{3}    & China            & 675 (6.90\%)   & Canada           & 1118 (6.49\%)  & Germany           & 19 (6.49\%)  \\ \hline
{4}    & France           & 663 (6.77\%)   & Russia           & 846 (4.91\%)   & China           & 17 (5.21\%)   \\ \hline
{5}    & Netherlands      & 475 (4.85\%)   & Germany          & 783 (4.54\%)   & Netherlands          &   17 (5.21\%)             \\ \hline
{6}    & Canada           & 369 (3.77\%)   & United Kingdom   & 559 (3.24\%)   &  United Kingdom    &  16 (4.91\%)             \\ \hline
{7}    & {---}               & 368 (3.76\%)   & Netherlands      & 470 (2.73\%)   & France   & 11 (3.42\%)   \\ \hline
{8}    & United Kingdom   & 307 (3.14\%)   & South Korea      & 429 (2.49\%)   & Brazil      & 11 (3.42\%)   \\ \hline
{9}    & Russia           & 296 (3.02\%)   & France           & 399 (2.31\%)   & Canada      & 11 (3.42\%)   \\ \hline
{10}   & Japan            & 219 (2.24\%)   & Japan            & 279 (1.62\%)   & Hong Kong           & 11 (3.42\%)   \\ \hline
\end{tabular}}
\end{table*}

\subsubsection{Case studies} A 51\% attack is not beyond the realm of possibilities. In July 2014, a Bitcoin mining pool ``GHash.IO'' acquired over 51\% of the hash rate for one day \cite{bastiaan2015preventing}, which raised many concerns in the press and media about Bitcoin and its vulnerabilities, and shed light on the general problem in \bc-based systems. Although no malicious activity was carried out, ``GHash.IO'' later shrunk in size when miners left its pool and eventually closed in October 2016. In August 2016, a group of attackers, known as ``51 crew'', hijacked two Ethereum Blockchains, namely Krypton and Shift, and managed to hijack 21,465 Kryptons worth of digital currency by double-spending. In May 2018, a group of malicious miners acquired 51\% hash rate in Bitcoin Gold and stole \$18 million USD worth of cryptocurrency \cite{Roberts_18}. In June 2018, four other notable Blockchain-based cryptocurrencies were also attacked; namely Monacoin, Zencash, Verge, and Litecoin Cash.

\subsection{Network Attacks}\label{sec:dns}

Blockchain applications are decentralized and use peer-to-peer network architecture as the medium of communication between the network entities. In this section, we will look into the attacks associated with the peer-to-peer network and we will use Bitcoin network as our example to provide details of these attacks. The attacks associated to the Blockchain network include among others, the DNS attacks, spatial partitioning, and Eclipse attacks. For each of these attacks, the goal of the attacker is to isolate users and miners from the real network, limit their access to the network resources, or create partition in the network and enforce conflicting rules among the peers.

\subsubsection{DNS Attacks} When a node joins the Bitcoin network for the first time, it is not aware of the active peers in the network. To discover the active peers (identified by their IP addresses) in the network, a bootstrapping mechanism is required. The Domain Name System (DNS) can be used as a bootstrapping mechanism, and DNS seeds are queried by nodes upon joining the network to obtain further information about other active peers. The initial DNS query returns one or more DNS A records along with their corresponding IP addresses of peers that are accepting incoming connections. Once the new node establishes connections to the peers, it can send {\em addr} command with port numbers to establish connections with other peers.

\begin{figure}[t]
\begin{center}
\includegraphics[ width=0.45\textwidth]{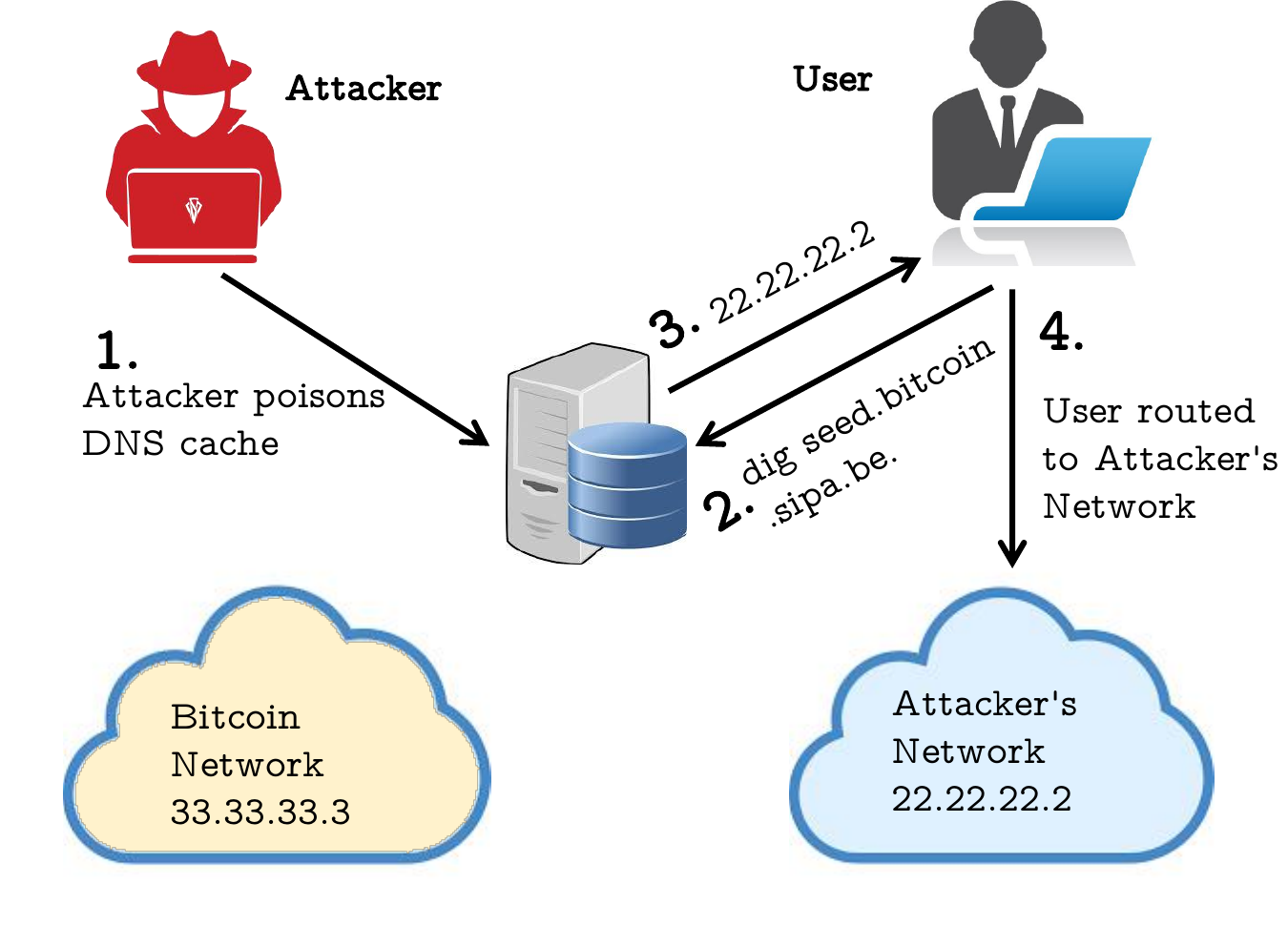}
\caption{DNS resolution attack on Bitcoin. The attacker poisons DNS cache and modifies the data. When a user queries the server to obtain IP addresses of peers who are accepting connections, he is routed to attacker's network. The attacker can game the user by feeding him fake blocks and transactions.}\label{fig:dns}
\end{center}
\end{figure}

It has been mentioned in the developer's guide of \bc systems \cite{developer} that the DNS opens a wide attack surface to the \bc networks in general. Namely, the DNS resolution is vulnerable to man-in-the-middle attacks (at the resolver side), cache poisoning, and stale records, among many others. For this attack, an adversary can either inject an invalid list of seeder nodes in the open source Blockchain software, or poison DNS cache at the resolver. By default, the Blockchain software client has a list of seeders that allow the network discovery. If the attacker injects a fake list of seeders, the user will be compromised. As a result, the adversary can potentially isolate Blockchain peers and lead them to a counterfeit network. In~\autoref{fig:dns}, we illustrate how a DNS attack can be carried out by poisoning DNS cache. A node in Bitcoin network has an IP address of 33.33.33.3 (for illustration purpose only) while the attacker's node in a counterfeit network has an IP address 22.22.22.2. The attacker poisons the DNS cache to lure the user into the counterfeit network. The user makes the DNS query {\em dig seed.bitcoin.sipa.be.} and instead of responding with 33.33.33.3, the DNS resolver returns 22.22.22.2. As a result, the user connects to malicious nodes in the counterfeit network and malicious nodes may feed false blocks to the user. For more on DNS security, we refer to the work in~\cite{KangSM16}. 
\begin{figure}[t]
\begin{center}
\includegraphics[ width=0.48\textwidth]{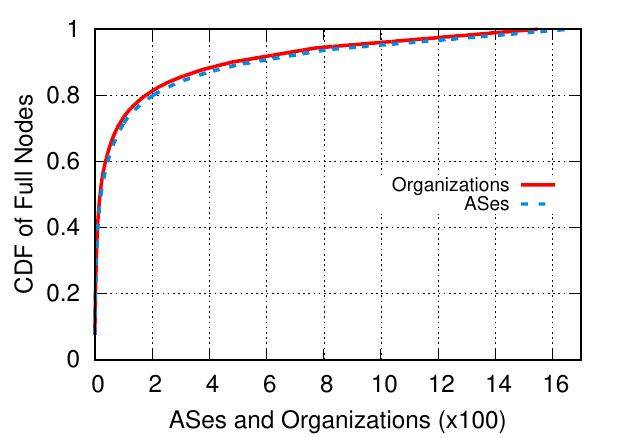}
\caption{Distribution of full nodes in Bitcoin across ASes and ISPs (organizations). Notice that less than 50 ASes and ISPs host more than 50\% of nodes showing that the network is centralized and vulnerable to BGP attacks.}
\label{fig:cdf}
\end{center}
\end{figure}

\subsubsection{BGP hijacks and spatial partitioning} There are two types of nodes in most Blockchain applications, namely full nodes and lightweight nodes. Full nodes are the actual participants in the network responsible for relaying blocks and transactions and maintaining an updated copy of the Blockchain. Lightweight nodes do not maintain a Blockchain and only use the services of full nodes to get access to the network. Since lightweight nodes draw their view of the Blockchain from the full nodes, when a full node is compromised all of its associated lightweight nodes are also compromised. Full nodes in a Blockchain network are spatially distributed across the Internet. In \autoref{tab:spa}, we show the spatial spread of full nodes in three major \bc systems (\cc). In each system, a majority of the nodes is located in United States, Germany, China, and Russia. The flow of traffic on the Internet is controlled by Internet Service Providers (ISPs), which own one or more Autonomous Systems (ASes), responsible for handling traffic routing. \cite{Gao01},\cite{KumarK14}.   

\begin{figure*}[t]
\begin{center}
\includegraphics[ width=0.78\textwidth]{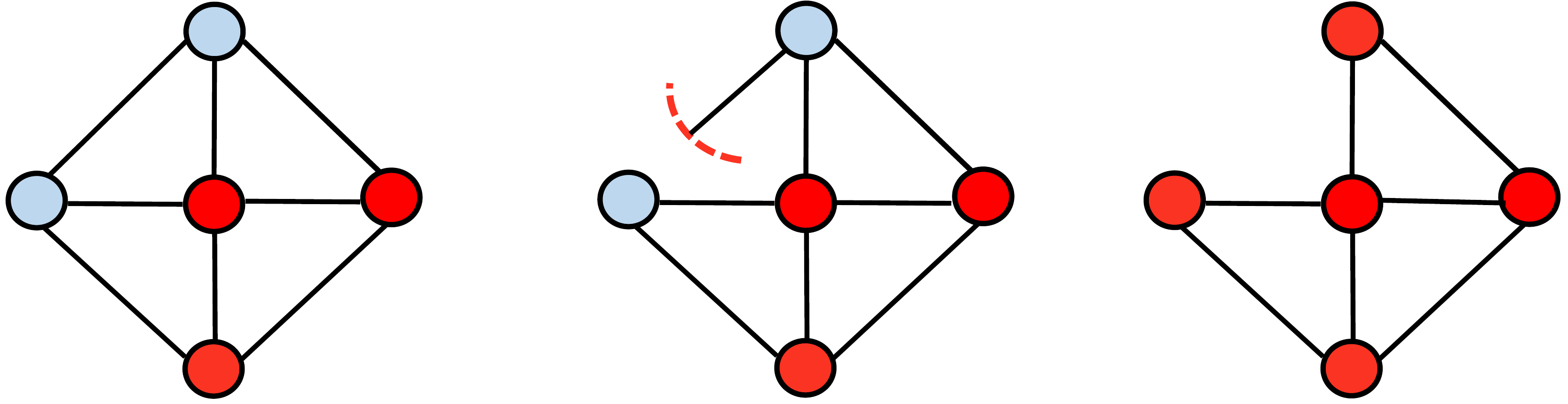}
\caption{Eclipse attack on a \cc network. Here, blue nodes represent the honest nodes following the true state of Blockchain while the red nodes represent the malicious nodes that form a cluster around the blue nodes. If the connection between the honest nodes is compromised, the malicious nodes may feed fake blocks to the honest nodes and partition them from the network. As a result, the honest nodes end up having the wrong view of the Blockchain.}
\label{fig:eclipse}
\end{center}
\end{figure*}

Spatial concentration of nodes within an AS or an ISP makes them vulnerable to routing attacks such as BGP hijacking. An adversarial AS can hijack the traffic for a target AS that hosts a majority of the Blockchain application nodes. This can disrupt the flow of valuable information, including transactions and blocks, to the nodes being hosted by the target AS. When the victim nodes are miners or mining pools, the attacker can substantially reduce the hash rate of the Blockchain application, thereby affecting the system activities. In a mining pool, the miners communicate using stratum overlay protocol. The stratum servers act as a dropzone where miners submit their PoW. Stratum servers have a public IP address that makes them vulnerable to routing attacks and flood attacks. Apostolaki~\etal\cite{apostolaki2017hijacking} studied that by hijacking fewer than 100 border gateway protocol (BGP) prefixes in Bitcoin, an attacker can isolate up to 50\% of the network's hash rate. They further explored that 60\% of all Bitcoin traffic traverses only three internet service providers (ISPs). Every month, over a 100 Bitcoin nodes suffer from routing attacks and BGP hijacks. Furthermore, they estimated that the routing attacks can delay block propagation by up to 20 minutes. As mentioned in~\autoref{sec:stale}, the average block computation time in Bitcoin is 10 minutes. Therefore, the routing attacks can delay the propagation of two or more blocks to a group of nodes. Such delays increase the likelihood of other attacks including Blockchain fork, consensus delay, and double-spending.

To verify their results and further analyze the spatial vulnerability of Bitcoin network, we replicated their study and noticed that Bitcoin network has further centralized with respect to ASes and ISPs. We crawled data from ``Bitnodes'', an online service that maintains information related to full nodes in Bitcoin \cite{bitnodes_18}. In \autoref{fig:cdf}, we plot the CDF of the spatial distribution of full nodes across ASes and ISPs in the world. In \autoref{tab:mp}, we show the distribution of mining pools across ASes and ISPs in Bitcoin. Notice that 60\% of the hash rate is solely intercepted by {\em AliBaba}. Our results show that compared to the prior work by Apostolaki~\etal\cite{apostolaki2017hijacking}, the Bitcoin network has further centralized and become more vulnerable to routing attacks.

\begin{table}[t]
\centering
\caption{Top 5 mining pools per hash rate, ASes, and organizations. 65.7\% of mining data goes through only three organizations. Alibaba alone has a view of at least 60\% of the mining data. We exclude the remaining 12 mining pools from the study as their total contribution to hash rate is minimal.   }
\label{tab:mp}
\begin{tabular}{|l|c|c|c|}
\hline
\multicolumn{1}{|l|}{\textbf{Mining Pool}} & \textbf{H. Rate \% } & \textbf{ASes} & \textbf{ISP} \\ \hline
{\multirow{2}{*}{BTC.com }}                                & {\multirow{2}{*}{25\%  }}                 & AS37963       & Alibaba       \\ \cline{3-4} 
                                           &                       & AS45102       & AliBaba        \\ \hline
\multicolumn{1}{|l|}{Antpool}              & 12.4\%                & AS45102       & AliBaba        \\ \hline
\multicolumn{1}{|l|}{ViaBTC}               & 11.7\%                & AS45102       & AliBaba        \\ \hline
\multicolumn{1}{|l|}{BTC.TOP}              & 10.3\%                & AS45102       & AliBaba        \\ \hline
{\multirow{2}{*}{F2Pool}}                                    & {\multirow{2}{*}{6.3\%}}                 & AS45102       & AliBaba        \\ \cline{3-4} 
                                           &                       & AS58563       & Chinanet        \\ \cline{1-4}

{\multirow{1}{*}{12 others}}                                    & {\multirow{1}{*}{34.3\%}}                 & ---      & ---       \\ \cline{1-4} 

\end{tabular}
\end{table}

\BfPara{Case studies} Over the last few years, a number of BGP attacks have been launched against ASes that host mining pools or cryptocurrency exchanges. In 2014, a malicious ISP in Canada announced BGP prefixes belonging to major ISPs including {\em Amazon, OVH, Digital Ocean, LeaseWeb,} and {\em Alibaba}, and intercepted the traffic routed to mining pools. As a result, the attacker made a fortune of 83,000 USD. In April 2018, BGP attacks were launched against {\em MyEtherWallet.com}, an open source web application used for exchanging Ethereum tokens online. Attackers managed to steal 152,000 USD from the web application \cite{greenberg_2017}.

\subsubsection{Eclipse Attacks} Blockchain's peer-to-peer system is also vulnerable to a form of attack known as the eclipse attack \cite{HelimanKZG15,MarcusHG18,NayakKMS16}, in which a group of malicious nodes isolates its neighboring nodes using IP addresses, thereby compromising their incoming and outgoing traffic. For example in Bitcoin, a node can actively connect to all the other nodes in the network, forming a node cluster. In the node cluster, every peer is aware of the IP address of all other peers. With sufficient compromised nodes in a cluster, the attacker can isolate honest nodes and change their Blockchain view. He can control their incoming and outgoing traffic and feed them with fake information regarding Blockchain and transactions. 

In \autoref{fig:eclipse}, we illustrate this attack procedure. As long as the honest node maintains a connection with one other honest node, it is likely to receive the correct information to maintain the true state of the Blockchain. However, when the connection between the honest nodes is compromised, they will get surrounded by malicious nodes and become vulnerable to the eclipse attack. When such nodes are fed with fake transactions and blocks, they eventually develop the wrong view of the state of Blockchain and become part of the malicious node cluster. Furthermore, if another honest node establishes a connection with the malicious node cluster, it is also exposed to the same vulnerability which leads to the cascade effect of propagation of fake transactions and blocks.

\subsection{Distributed Denial of Service Attacks} \label{sec:DDOS}
One of the most common attacks on online services is the distributed denial-of-service (DDoS) attack~\cite{WangMC17}. Blockchain technology, and despite being a peer-to-peer system, is still prone to DDoS attacks. Blockchain-based applications, such as Bitcoin and Ethereum, have repeatedly suffered from these attacks \cite{vasek2014empirical,DDOS1,cimpanu_2017, ethereum_blog_2016, cryptocoinsnews_2016}.  DDoS attacks manifest themselves in a number of ways, depending upon the application nature, network architecture, and peers behavior. For example, in the Bitcoin network, the 51\% attack can lead to denial-of-service. Specifically, if a group of miners acquire a significant hashing power, they can prevent other miners from adding their mined blocks to the Blockchain, invalidate ongoing transactions, and cause service failure in the network. Intentional forks; forks that are the result of malicious behavior; can turn into hard forks, resulting in similar outcomes of denial-of-service.

\subsubsection{Stress testing} Another possibility for the attack is due to the limited number of transactions per block a Blockchain application can process in a given time. For example, on average, it takes the Bitcoin network 10 minutes to mine a block, which has a maximum size of 1MB. 
Although the size of transactions in \bc varies, the average size of a transaction in \bc is approximately 500 bytes, allowing approximately 2,000 transactions per block on average---the maximum number of transactions added to a block in \bc is reported to be 2,210~\cite{block-explorer}. 
Furthermore, the average time needed to mine a block, based on the predefined difficulty, is approximately 10 minutes. As such, for all current transactions in the network to be successfully included in the Blockchain, their number may not exceed 200 transactions per minute. Taking that into account, and the fact that each transaction requires a minimum of two peers (identified by two different public identifiers) to be involved in a transaction, the total active peers served by the network per minute (\textit{i.e}. where a block containing their transaction will be mined) will not exceed 200 peers. Given these constraints, the throughput of Bitcoin is 3-7 transactions per second. Throughput of Bitcoin is low compared to mainstream payment processors such as Visa Credit that can verify up to 2,000 transactions per second. 

An adversary may exploit the aforementioned operational reality of the Bitcoin system by introducing Sybil identities; the same adversary also may control multiple wallets. Furthermore, using those identities, the adversary may issue several dust transactions (\eg, 0.001 BTC per transaction) between the various Sybil identities under his control. By introducing a large number of transactions of small value over a short period of time, the network will be congested by creating blocks containing those transactions, and service to legitimate users in the network will be denied. As a result of this congestion the adversary may as well launch other attacks; \eg, double-spending of tokens not mined due to the congestion. 

One may argue that miners may choose which transactions to include in a block. However, this is discouraged by design in Bitcoin, as outlined by Satoshi. Blocks today even include transactions of value as low as 0.0001 BTC, which makes flooding the network with low-value transactions possible. 

\begin{figure}[t]
\begin{center}
\includegraphics[ width=0.48\textwidth]{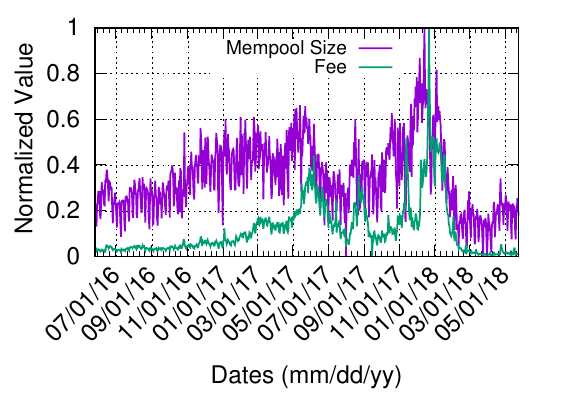}
\caption{Bitcoin mempool size and average fee paid to the miner. It can be observed from the figure that as the mempool size is increases, the fee paid to the miners also increases. }
\label{fig:mempool}
\end{center}
\end{figure}

\subsubsection{Mempool flooding} Another form of DDoS attack is carried out at the memory pools (mempools) of the cryptocurrencies to increase the mining fee. As outlined in~\autoref{fig:overview}, mempools act as a cache of unconfirmed transactions. Although the block size is limited in the cryptocurrencies, the mempool size has no size limit. However, users estimate the size of mempools to prioritize their transactions. If there are more transactions in the mempool, then the competition for mining becomes high. To prioritize their transactions, users start paying more mining fees as an incentive for miners. Saad \etal \cite{saad2018poster}, identified a low cost DDoS attack on Blockchain applications in which the adversary along with Sybil nodes may flood the mempools with unconfirmed transactions. Such an attack creates panic among the legitimate users who are tempted to pay higher mining fee to prioritize their transactions while the attacker's transactions do not get mined. As a result, the attacker launches a DDoS attack.

\subsubsection{Case Studies} In Bitcoin, malicious users have been flooding the mempool with dust transactions to make legitimate users pay higher mining fees. On November 11, 2017, the Bitcoin mempool size exceeded 115k unconfirmed transactions, resulting in \$700 million USD worth of transaction stall \cite{cryptocoinsnews_2017}. In June 2018, again the mempool was attacked with 4,500 unconfirmed spam transactions which increased the mempool size by 45MB. The increased size led to a spike in the mining fee and legitimate users were propelled to pay higher fee to get their transactions mined \cite{mempool_2018}. In~\autoref{fig:mempool}, we plot the mempool size and the mining fee of Bitcoin over the last two years. We use min-max normalization to scale the data points.

\subsubsection{DDoS Attacks in Private Blockchains} In PBFT-based private Blockchains, a DDoS attack can be launched if the adversary controls $\approx$33\% replicas~\cite{CastroL02}.  In the private Blockchains, the size of the network is known to the participating nodes, which allows the adversary to calculate the number of sybil nodes he needs to introduce in the network for an attack. Assuming that the adversary controls $f$ sybil nodes such that the total network size is $n < 3f+1$, then the attacker will be able to launch a DDoS attack to stop the verification process. For each transaction sent by the primary, the sybil replicas will not reply with their approvals. Since the primary will need approvals from at least $3f+1$ replicas, it will not be able to process any transaction, and the system activities will be halted leading to a DDoS attack. 

In public Blockchains, launching such an attack can be costly. The adversary needs to have either the majority of the total hash rate, the majority of stake, or control over 50\% network peers. Considering that public Blockchain applications such as Bitcoin have more than 10,000 active full nodes~\cite{bitnodes_18}, it is infeasible for the adversary to launch a successful attack. On the other hand, in private Blockchains, the network size does not grow beyond a few hundred nodes, whereby the adversary needs to control only $33\%$ replicas or just the primary replica, making the attack on private Blockchains more feasible. 

\subsection{Block Withholding Attacks} \label{sec:faw}
The peer-to-peer network of cryptocurrencies can be exploited to create conflicting views about the Blockchain. Malicious nodes can intentionally mask, forge, or withhold important information that needs to be relayed across the network. Some of the known attacks of this nature are ``The Finney Attack'' and ``Block Withholding Attack'' \cite{tosh2017security}. 

\subsubsection{The Finney attack} The Finney attack is a variant of the double-spending attack in which a miner delays block propagation to double-spend his transaction \cite{mark_2017,stackexchange_finney}. The miner generates a transaction, computes a block, and chooses not to relay the block. In the meantime, he generates a duplicate of his previous transaction and sends it to a recipient. After the recipient accepts the transaction and delivers the product, the miner publishes his previous block with the original transaction in it. Therefore, the previous transaction sent to the recipient becomes invalid and the miner successfully double-spends transaction. 

The Finney attack has low success probability due to short block intervals and time sensitive attack procedure. The block time in Bitcoin and Ethereum are 10 minutes and 15 seconds. If an attacker attempts to launch this attack on Ethereum, it is unlikely that he will be able to 1) generate a double spent transaction, 2) trick an optimistic receiver, 3) receive product before confirmation, and 4) publish a block before any other miner, within 15 seconds. Since the attack procedure is more time consuming than the block interval time, Finney attack is highly infeasible and as such, no case of Finney attack has yet been reported on any \cc.

\subsubsection{Classical block withholding attack} The block withholding attack is launched against decentralized mining pools with intent to harm the pool operator by withholding a valid PoW. \cite{LuuSPSH15,stackexchange_bwa}. In decentralized mining pools, all participants consume electricity and CPU power to find a nonce whose value of a hash with the block is less than the target threshold. Once the valid solution is found, all participants are rewarded based on their aggregate effort put towards the computation of the solution. Since nonce finding is a lottery-based system, therefore, miners with less hash power may come up with a valid solution before other miners with a higher hash rate. In  the block withholding attack, a compromised miner in the pool finds the proof-of-work and chooses not to disclose it to the pool operator. Unaware of the compromised miner, the rest of the miners in the pool waste their resources to find the nonce and eventually lose the race. The malicious miner then can collude with other mining pools and share the PoW with them for a higher reward, or even publish the block independently with a different identity. Due to this unfair behavior of one miner in the pool, the entire pool is deprived of block rewards. 

Another form of withholding attack is possible when two mining pools intentionally try to fork the Blockchain to create a network partition \cite{kwon2017selfish}. For instance let there be two mining pools in a \cc, namely $Mp_A$ and $Mp_B$,  and $Mp_A$ computes a valid block but decides not to publish it. $Mp_A$ waits for $Mp_B$ to compute and publish the block. As soon as $Mp_B$ releases its block, $Mp_A$ also releases its block and resulting in two valid blocks in the network. This will fork the Blockchain and nodes in the network will have a consensus disagreement upon receiving two valid blocks. Although this attack may partition the network, it may also cause loss to both mining pools. Therefore, no such attack has been reported in any Blockchain application so far.

\subsubsection{Fork after withholding attack} Another form of withholding attack is known as the fork after withholding (FAW) attack. Introduced by Kwon~\etal \cite{kwon2017selfish}, FAW is always more rewarding than block withholding attacks. In the following, we outline the attack procedure of FAW.

\begin{enumerate*}
    \item A malicious miner joins two mining pools $Mp_A$ and $Mp_B$ respectively. 
    \item The miner computes a valid PoW in mining pool $Mp_A$.
    \item He withholds the solution and only publishes the block once $Mp_B$ also publishes the block. 
    \item The network selects one block among the two.
    \item The malicious miner gets rewarded either way.
\end{enumerate*}

Kwon~\etal \cite{kwon2017selfish} also show that if the FAW attack is launched by two or more mining pools against each other, then the bigger mining pool will always win in the race condition. Therefore, the FAW attack is always more profitable than selfish mining and block withholding. 

\subsubsection{Block Withholding in Private Blockchains} In private Blockchains, the primary replica can launch a block withholding attack after receiving a confirmations from other replicas. Private Blockchains work under the assumption that the primary will faithfully execute the protocol. Moreover, the adversarial model assumes that the attacker controls a subset of faulty replicas among all the other replicas. However, if the adversary also controls the primary replica, he can withhold blocks and transactions from all other replicas. As shown in \autoref{fig:pbft}, the primary receives a transaction request from the client and sends the transaction to other replicas to obtain their signatures. Finally, it computes the block when a sufficient number of transactions are processed. However, if the primary gets compromised, he may: 
\begin{enumerate*}
    \item withhold a transaction issued by the client and abort the verification process,
    \item delay the verification process by sending the transaction to fewer replicas, 
    \item receive the signatures and discard them, 
    \item compute a block and withhold it from the rest of the network.
\end{enumerate*} 
In each case, the primary can launch a withholding attack to compromise the system and delay the transaction processing. 

\begin{figure}[t]
\begin{center}
\includegraphics[ width=0.4\textwidth]{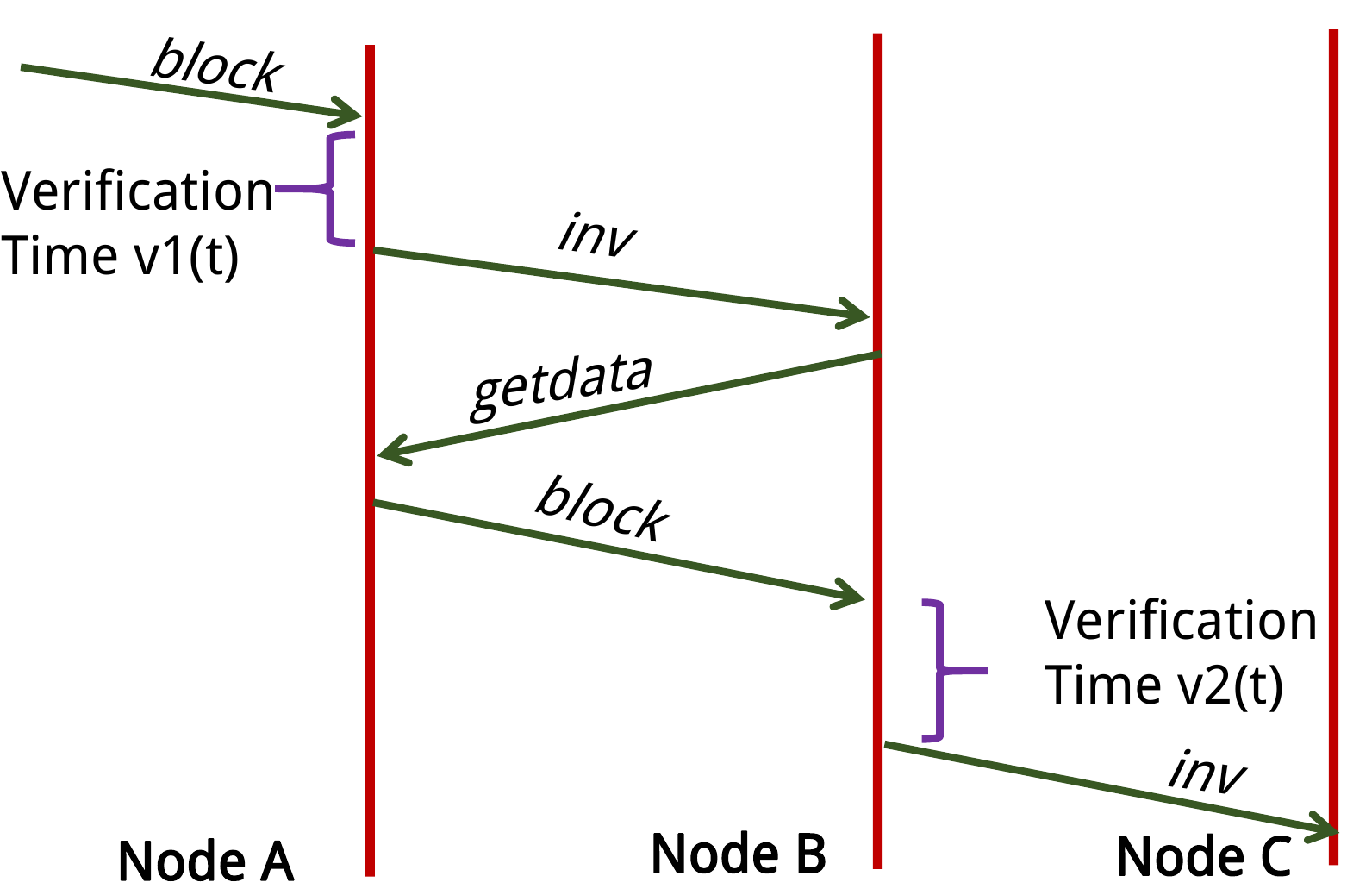}

\caption{Block propagation between nodes A, B, and C. Notice that the maximum time is consumed in block verification, $v1(t)$ and $v2(t)$. Consensus delay can be caused by propagating false or stale blocks in the network.  }
\label{fig:delay}
\end{center}
\end{figure}

\subsection{Consensus Delay} \label{sec:consensus} 
Another attack associated with the peer-to-peer nature architecture is the consensus delay, noticed by Geravis \etal \cite{GervaisRKC15}. In this attack, an attacker may inject false blocks to add latency or prevent peers from reaching consensus about the state of the Blockchain. In~\autoref{fig:delay}, we illustrate the delays incurred during block propagation in Bitcoin. When node A receives a block, it authenticates the block and sends an \textit{inv} message to its neighbors including node B. If node B does not have the block, it sends \textit{getdata} message back to A. Upon receiving the \textit{getdada} message from B, A sends the block to B. Once B has the block, it also authenticates the block and sends an \textit{inv} message to its neighbors. As~\autoref{fig:delay} shows, the maximum delay is incurred during authenticity check ($v1(t)$ and $v2(t)$). The other delays include transmission delays and propagation delays of messages and block. Transmission delays are subject to the size of block and messages while the propagation delays depend upon the bandwidth of the link between the nodes. 

In such conditions, intentional delays can be introduced in the network by propagating stale blocks or double-spent transactions. The nodes which are not aware of stale blocks will respond with \textit{getdata} messages and upon receiving the block, they will waste time in its verification. If an attacker controls a set of sybil nodes in the node cluster~\autoref{sec:dns}, it can add significant delays among legitimate nodes in that cluster. The problem is further exacerbated for time-critical applications such as Blockchain-based peer-to-peer gaming, where resolution needs to be achieved within short time. 

In PBFT-based private Blockchains, an adversary can also cause consensus delay by using sybil nodes. As shown in \autoref{fig:pbft}, a major component of transaction processing is the exchange of messages and signatures among participating replicas. Especially, in the {\em prepare} and {\em commit} phase, each replica sends its signatures to every other replica. As outlined in \autoref{sec:DDOS}, if the adversary controls more than 33\% of replicas, he can launch a DoS attack. On the other hand, if the adversary controls fewer replicas, he may still be able to cause consensus delay in transaction processing~\cite{CastroL99,XuLLLG18}. The sybil nodes can also send bogus signatures to the other replicas during the prepare phase and the commit phase. Since each replica is then required to verify signatures, therefore, bogus signatures will cause additional verification overhead. If the sybils continue to send such signatures, they can stall the completion of the commit phase and eventually cause a delay in the reply phase. As a result, the primary will not receive the required number of approvals for the transaction verification. This will cause consensus delay and reduce the throughput of the application.

\subsection{Timejacking Attacks} \label{tj}
In \bc systems, such as Bitcoin, full nodes maintain an internal counter that denotes the network time. The network time is obtained by receiving a version message {\em ver}  from neighboring peers and calculating its median during the bootstraping phase. If the median time of all the neighboring peers exceeds 70 minutes, the network time counter is automatically reverted to the system time of the node. This creates an attack opportunity for malicious nodes which may connect to the target node as shown in \autoref{fig:eclipse}. In such case, the attacker can feed varying timestamps with median value exceeding 70 minutes. Furthermore, in Bitcoin, for example, a node rejects a block if its timestamp exceeds the network time by 120 minutes. An attacker can compute a new block and set its timestamp ahead of network's timestamp by 50 minutes. The attacker then, along with Sybil nodes, can slow the network time of a target node by launching a timejacking attack against it. As a result, the difference between the block time and the target node's counter will exceed 120 minutes. As a result, if the target node is presented with the block, it will reject it and all the subsequent blocks. The target node eventually gets isolated from the activities of the main network.

\subsection{Countering Peer-to-Peer Attacks}\label{sec:cp2p}
Prior research has been conducted to address the problem of selfish mining, and researchers have suggested several possible solutions \cite{sapirshtein2016optimal,eyal2014majority,heilman2014one,courtois2014subversive}. Solat and Potop-Butucaru \cite{solat2016zeroblock} proposed a ``lifetime'' for blocks that prevents \textit{block withholding} by selfish miners. If the expected lifetime of a block expires (calculated by the honest miners), it is rejected by the network. Heilman \cite{heilman2014one} impedes the profitability of selfish miners by introducing a defense scheme called ``Freshness Preferred.'' Heilman \cite{heilman2014one} builds on top of the previous work by Eyal and Sirer \cite{eyal2014majority} by adding unforgeable timestamps to blocks and prefers blocks with more recent timestamps compared to older ones. His work reduces the incentive for selfish miners to withhold their blocks for long periods of time. Eyal \cite{eyal2015miner} modeled a game between two mining pools carrying out  \textit{block withholding} and discovered \textit{miner's dilemma}, where both mining pools suffer a loss in equilibrium. 

Majority attacks have also been widely discussed with countermeasures proposed to overcome a monopoly in Blockchain networks. Miller \etal\cite{MillerKKS15} proposed changes to the PoW puzzle in Bitcoin in order to restrict coalitions of mining pools for majority attacks. Their proposed design incorporates {\em nonoutsourceable} puzzles in PoW, in which mining pools that outsource their mining work risk losing mining rewards. Saad~\etal \cite{SaadLCA18} leveraged the expected transaction confirmation height and the block publishing height
to detect selfish mining behavior in PoW-based Blockchains. Using the relationship between the two features, they created a ``truth state'' for each published block in order to distinguish between a legitimate block and a selfishly mined block. Also addressing the 51\% attack, Bastiaan \cite{bastiaan2015preventing} introduced the concept of ``two phase proof-of-work'' (2P-PoW). 2P-PoW is a continuous-time Markov chain (CTMC) model that incorporates two challenges for miners to solve instead of one. The states of the CTMCs prevent the pool from increasing beyond an alarming size by shrinking incentive for miners in the pool. 2P-PoW prevents large pools from creating a hegemony by either outsourcing a major chunk of their hash rate or exposing the private keys of the pool operator.

Johnson~\etal \cite{johnson2014game} proposed a game-theoretic approach to address DDoS attacks against mining pools. Other countermeasures include putting a cap on the minimum amount in the transaction that a sender can have or increasing the block size to accommodate more transactions. Yet another approach is to reduce the difficulty in mining blocks so that more blocks can be mined with no transactions going to waste. Each of these propositions have their own caveats. 

Increasing the block size might not be sufficient, since a powerful adversary can still stress the network by generating dust transactions. On the other hand, reducing difficulty will reduce the block time but it will increase the number of orphaned blocks in the system and the overall size of the Blockchain. At the time of writing this paper, Bitcoin and Ethereum Blockchain size was recorded to be 162 GB and 450 GB, respectively \cite{statistabitcoin18}. Saad \etal \cite{saad2018poster} proposed fee-based and age-based countermeasures to prevent DDoS attack on Blockchain mempools. In their work, they shifted the transaction filtering process from the mining pools to the mempools. Their proposed countermeasures optimize the mempool size and raise the attack cost for the attacker while favoring legitimate users in the system. 

To prevent DNS-based attacks, extensive research has been carried out to equip the Blockchain systems with DNS attack defenses \cite{silva2009dnssec, peng2007survey, etheridge2002system}. Apostolaki~\etal \cite{apostolaki2017hijacking} proposed long- and short-term solutions for routing attacks. They propose routing-aware peer selections to maximize diversity of internet paths and limit the vantage points for attacks. They also proposed peer behavior monitoring to check abrupt disconnections and unusual latency in block delivery. 

Other solutions to prevent delay attacks include  end-to-end encryption for message propagation. Another possible approach to prevent spatial partitioning is the decentralized hosting of mining pools and full nodes over the Internet. As shown in \autoref{tab:spa}. 50\% of Ethereum nodes are located within two countries, which makes them vulnerable to a nation-state adversary. In order to prevent that, new nodes must be hosted on cloud services that have a higher geographical spread and network diversity. The dimensions we explored in this paper encourage additional research on Blockchain technology in the areas regarding DNS and DDoS attacks. 

To counter block withholding attacks~\cite{BagRS17,BagS16,courtois2014subversive,tosh2017security}, Schrijve \etal \cite{SchrijversBBR16} introduced an incentive-compatible reward scheme that discourages a malicious miner from carrying out withholding attacks against the targeted mining pool. Rosenfeld \cite{Rosenfeld11} introduced a Honeypot technique to lure rogue miners into a ``trap'', thereby catching the miner who withholds valid solutions. Bag and Sakurai \cite{BagS16} proposed additional incentives for finding a valid solution for a block in order to prevent mining collusion. Concurrent to their prior work, Bag \etal \cite{BagRS17} introduced a new scheme that blinds the miners in the pool from the current target to obfuscate their ability to distinguish between a partial and full PoW. Their proposed solution also binds the pool operator to fairly distribute the reward to the winning miner. 

The FAW attack can be countered by introducing timestampped beacons in the assignment given to the miners by the pool operators \cite{kwon2017selfish}. As a response to each assignment, the miners calculate the partial proof-of-work and send the response to the pool operator embedded with the beacon value. The beacon value is updated after a few seconds to catch a malicious miner if he withhold the valid solution and later propagates it in the network. However, the authors also noticed that this solution may not be practical in some situations and conclude that FAW attacks remain an open problem for the research community to address. To address the security issues in private Blockchains, several variants of PBFT protcol have been proposed. Those protocols try to increase the fault tolerance beyond 33\%~\cite{VeroneseCBLV13,DistlerCK16} and use hardware assistance to detect the behavior of faulty replicas~\cite{LiuLKA19}. The key challenge in private Blockchains is the high message complexity that restricts the scalability. As a result, in a small network, the adversary can easily compromise $33\%$ replicas. To address this issue, Liu \etal~\cite{LiuLKA19} proposed a scalable Byzantine consensus with hardware assisted secret sharing, which reduces the message complexity of PBFT to $O(n)$. This can be leveraged to construct large private Blockchain networks that can withstand various forms of attacks.

\begin{figure*}[htb]
\begin{center}
\includegraphics[ width=1\textwidth]{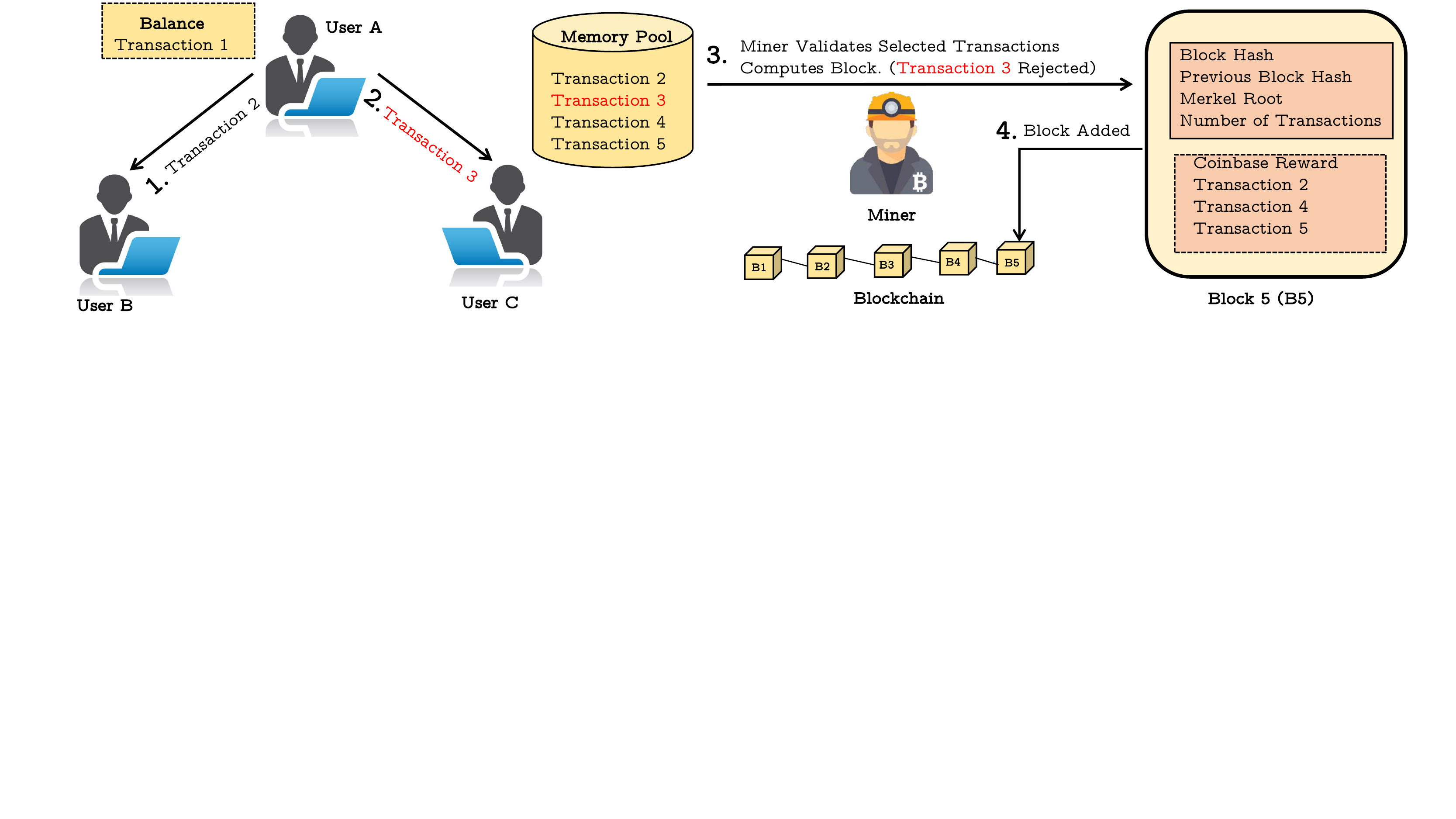}
\vspace{-62mm}
\caption{Double-spending attack carried out by User A. User A has Transaction 1 in his balance. Using that as an input, he generates Transaction 2 and sends it to user B. Then he generates Transaction 3 from the an already spent Transaction 1. When miner queries the mempool, he can either select Transaction 2 or Transaction 3. If Transaction 3 gets rejected, user C suffers the loss.   }
\label{fig:double-spending}
\end{center}
\vspace{-5mm}
\end{figure*}

\section{Application Oriented Attacks}\label{sec:application}
The Blockchain and associated peer-to-peer system are separate from the application services using them. Based on the nature of the Blockchain applications, they have their own vulnerabilities and attack surface. Therefore, we expect a significant number of attacks related to various applications, which we address in this section. Our analysis is primarily on applications such as cryptocurrencies and smart contracts.

\subsection{Blockchain Ingestion and Anonymity}\label{sec:ingestion}
Public Blockchains have a weak notion of anonymity, and they provide open data accessibility to the public. As such, the analysis of the public Blockchain can reveal useful information to an adversary. This process is known as Blockchain ingestion and it might not be desirable to the Blockchain application or its users. For example, a credit card company in the open market can use data analytics to delve into public information on the Blockchain and optimize its own schemes to compete with the digital currency. To demonstrate the potential exploitation of the public data, Fleder \etal \cite{fleder2015bitcoin} used graph analysis to create directed links between Blockchain data of Bitcoin and associated identities of the wallet users.

\BfPara{Mt. Gox incident} In 2013, two attackers exploited the public nature of Bitcoin Blockchain to carry out fraudulent transactions and create a fake demand of bitcoins at multiple exchanges. The main target of attack was Mt. Gox; the biggest Bitcoin exchange in Japan in 2013. The attackers frequently carried out a sequence of fraudulent transactions at Mt. Gox. Since the Blockchain is public, the rate of transactions was noted by other exchanges too and it was assumed as if the overall demand of the coins had increased. As a result, the price of Bitcoin increased from \$150 USD to \$1,000 USD towards the end of 2013. The trade carried out at Mt. Gox by the attackers was not backed by the real coins, which eventually led to the bankruptcy of the exchange.

\BfPara{Illegal Activities}  Anonymity in Blockchain-based cryptocurrencies provides lucrative opportunities for miscreants to carry out fraudulent activities. As such, cryptocurrencies have become a popular source of funds transfer for illicit activities associated with the {\em Deep Web} \cite{Janze17,Stokes12}. Since the use of fiat currency leaves traces on Blockchains that can be tracked by law enforcement, cryptocurrencies on the other hand, preserve the anonymity of the user. This is a key reason why various countries have banned the use of cryptocurrencies \cite{williams_2017}. 

Blockchains are tamper-proof, append-only, and decentralized; once a transaction is committed, it cannot be reversed. This has led to various irreversible scam activities online, where users are tricked into sending money through Bitcoin ATMs. Furthermore, the absence of a central authority makes it harder to claim fraud and expect reimbursement. Therefore, design constructs of Blockchain applications can be exploited to facilitate cyber crimes and online frauds.

\subsection{Double-Spending} \label{sec:double-spending}
In cryptocurrencies, double-spending refers to the use of a one-time transaction twice or multiple times. To illustrate double-spending with an example, consider the following scenario. In \cc operations, a transaction transfers the ownership of asset from a sender's address to the receiver's public address, and the value of the transaction is signed by the signer with a private key. Once the transaction is signed, it is broadcast to the network upon which the receiver validates the transaction. The validation at the recipient's end happens when the receiver looks up the unspent transaction output of the sender, verifies the sender's signature, and waits for the transaction to be mined into a valid block. The process takes a few minutes depending upon the size of the mempool, the throughput of the network, priority factor of the transaction, and the block computation time of the \cc. In \bc, the average time of block mining is 10 minutes.

In an environment of fast transactions~\cite{KarameAC12,KarameARGC15} or if a receiver is optimistic, he may release the product to the sender before the transaction gets mined into the Blockchain. As such, this gives the sender an opportunity to sign the same transaction and send it to another recipient. This behavior of signing the same transaction with a private key and sending it to two different receivers is known as double-spending. In double-spending there are two transactions derived from the same unspent transaction output of the sender, and only one of them gets incorporated into the Blockchain. In~\autoref{fig:double-spending}, we illustrate how a double-spending attack can be carried out in a \cc. Consensus delay in the network~\autoref{sec:consensus}, BGP attacks, flood attack on mempools, or the 51\% attack~\autoref{sec:51} can cause additional latency in the verification and propagation process, which increase the chances of an adversary to perform double-spending. In March 2013, due to a soft fork, a successful double-spending transaction worth \$10,000 USD was carried out in Bitcoin.
\begin{figure*}
\hfill

\begin{subfigure}[Google popularity index \label{fig:cj}]{\includegraphics[width=5.5cm]{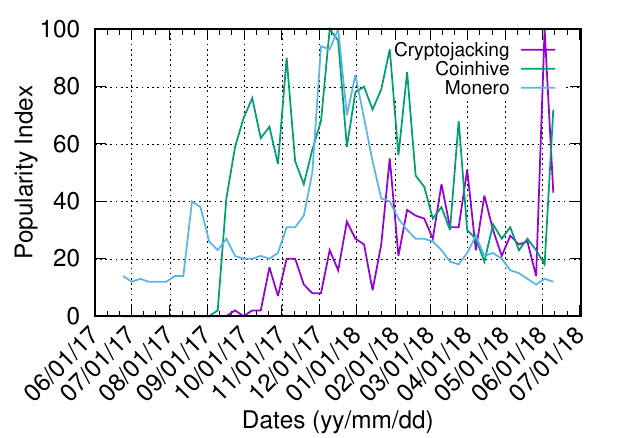}} 
\hfill
\end{subfigure}
\begin{subfigure}[ Percentage CPU usage by four \cj websites when JavaScript is enabled  \label{fig:dyn}]{\includegraphics[width=5.5cm]{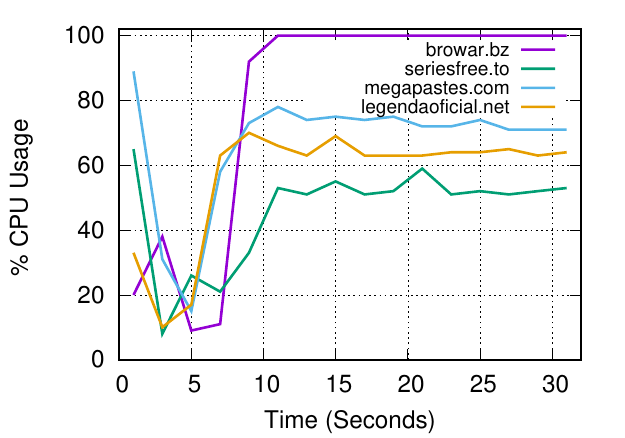}}
\hfill
\end{subfigure}
\begin{subfigure}[Percentage CPU usage by four \cj websites when JavaScript is disabled  \label{fig:maliciousX}]{\includegraphics[width=5.5cm]{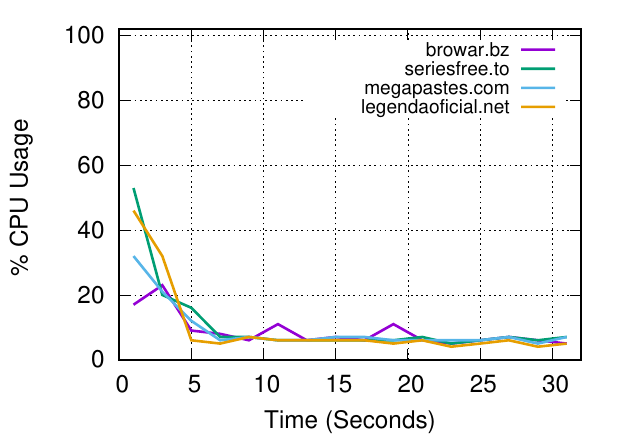}} 
\hfill
\end{subfigure}

\caption{(a) shows the Google search index for the terms ``Cryptojacking'', ``Coinhive'', and ``Monero.'' Notice that towards the end of 2017, there has been a rise in the Google search for the three terms which coincides with timing of large scale \cj attack. In (b) and (c) we show the effect of four \cj websites with an without JavaScript enabled. \Cj consumes high CPU power upto 100\% which can affect critical CPU operations.  } 
\label{fig:cryptojacking}
\end{figure*}

\subsection{Cryptojacking} \label{sec:cj}
\Cj is a form of an attack that is launched on web and cloud-based services to illegally perform PoW for Blockchain-based cryptocurrencies without consent~\cite{nadeau_2018,li_kyle_2018}. The most recent as well as the most prevalent form of \cj is the in-browser \cj, which turns websites into mining pools~\cite{SaadAA18}. PoW requires processor intensive mathematical calculations which usually involves finding a target hash value. As the aggregate hash rate of the \cc network increases, the associated difficulty to compute a block also increases. To meet the difficulty requirements, sophisticated hardware such as GPUs and ASIC chips \cite{goldbitcoin} are used by miners. Mining pools expand their hashing capability by inviting more miners to join their pool and purchasing expensive hardware with better computation capabilities. As a result, the mining process in major \bc systems becomes an expensive and competitive game that prevents small miners from mining blocks independently. 

\subsubsection{Cloud-based \Cj} To compensate for that, malicious miners have found a way of to expand their hash power by hijacking processors of remote devices for mining. This attack is known as the covert mining, or \cj attack, and it involves hijacking a target device to perform PoW calculations for the attacker. Initially, these attacks were launched against cloud service providers, where malicious users performed covert mining operations on virtual machines and exhausted cloud resources. This behavior was first noticed by Tahir \etal \cite{TahirHDAGZCB17}, where they also proposed countermeasures in the form of a software tool called ``MineGuard'' to effectively detect and stop covert mining operations in cloud.

\begin{figure*}[ht]
\centering
		\subfigure[Cryptojacking\label{fig:Cryptojacking}] {\includegraphics[width=0.3\textwidth]{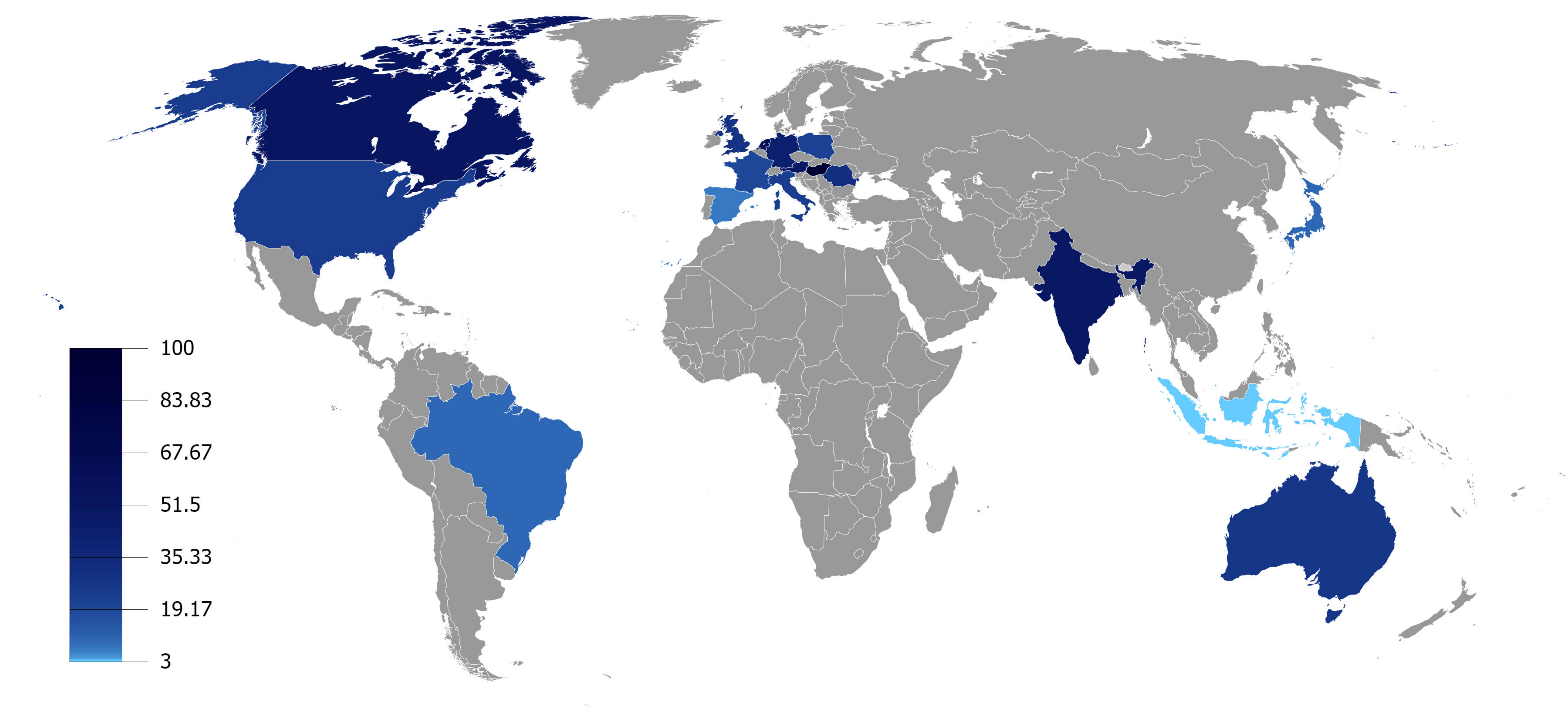}}
		\subfigure[Coinhive \label{fig:coinhive}] {\includegraphics[width=0.3\textwidth]{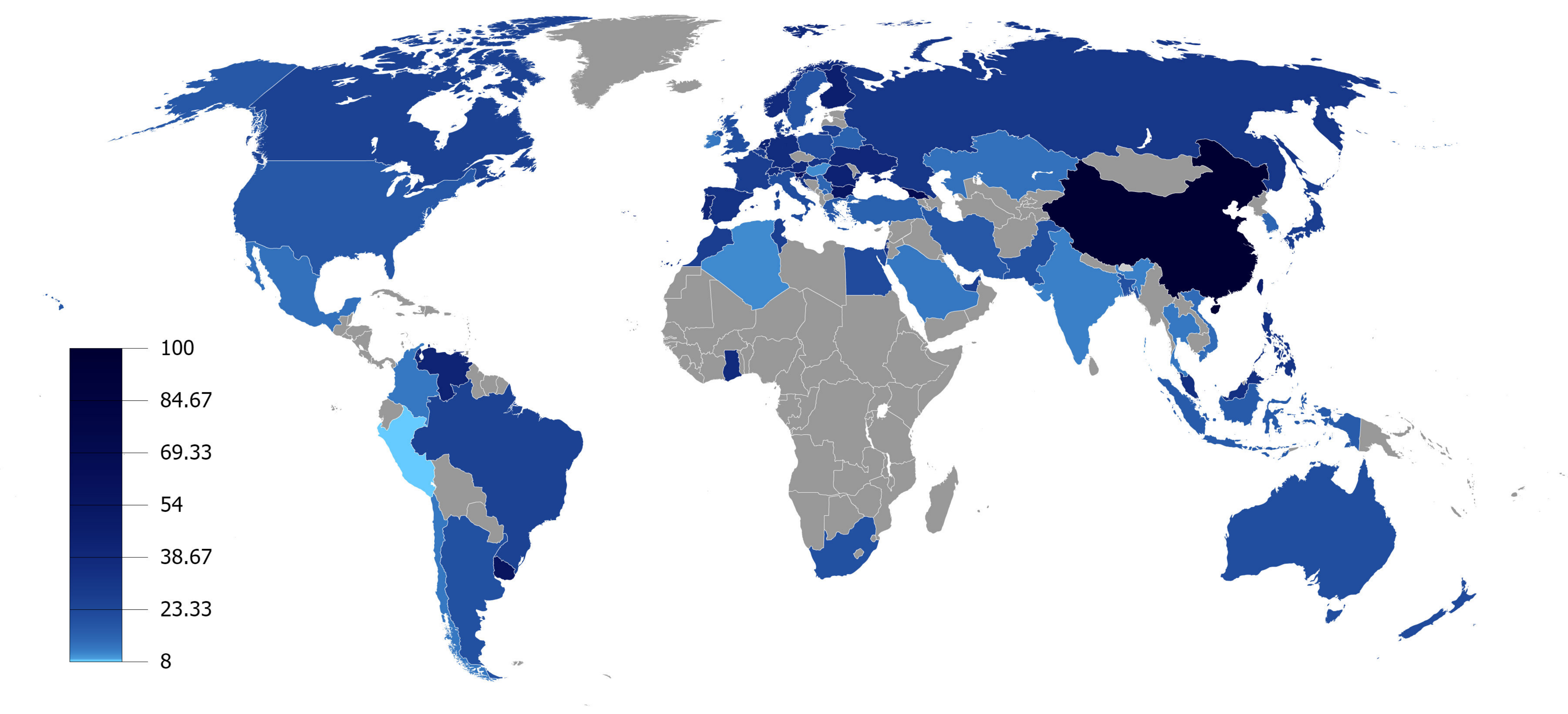}}
		\subfigure[Monero \label{fig:monero}] {\includegraphics[width=0.3\textwidth]{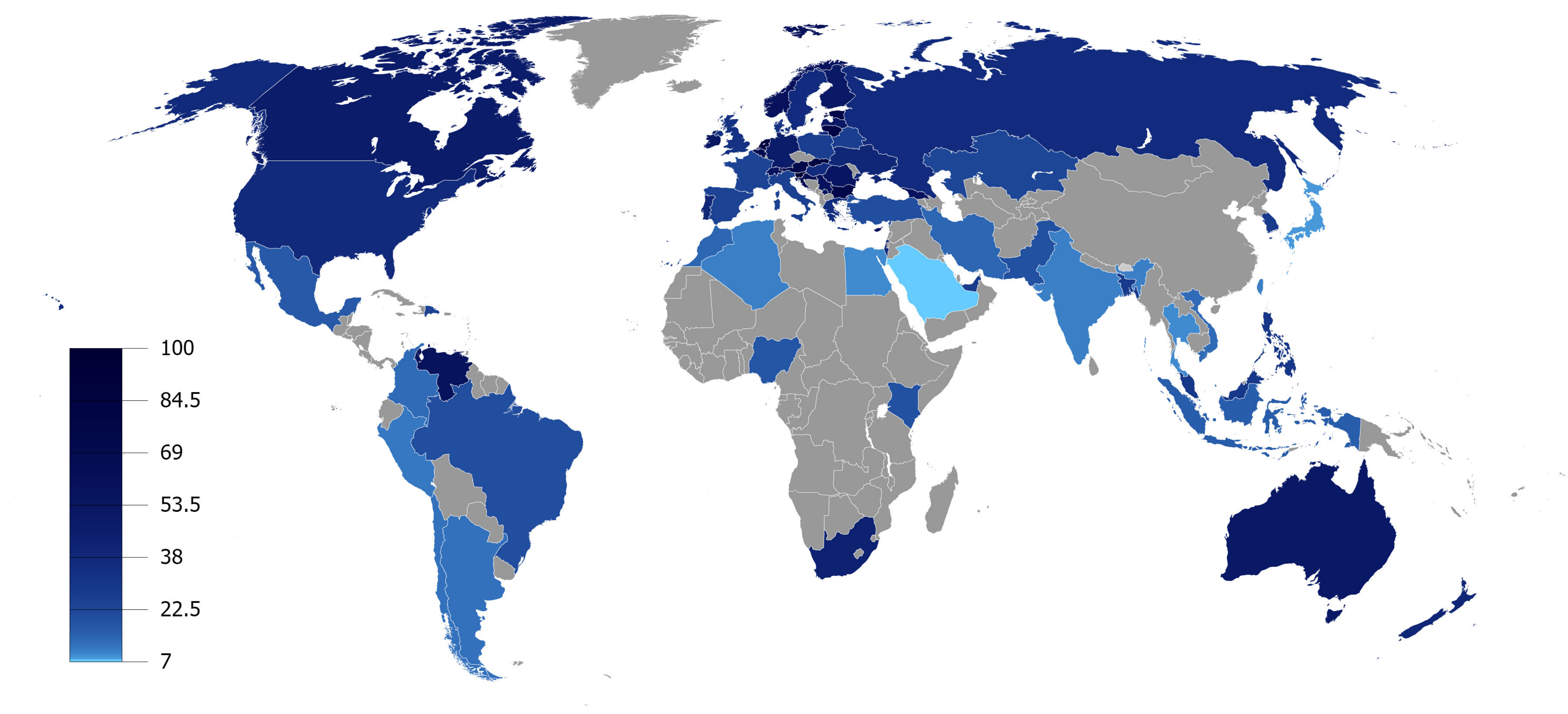}}
\caption{Heatmap of the global distribution of Google searches for each term. Notice that US is the most prevalent country in all three search results. Moreover there is more similarity in the search for Coinhive and Monero.}
\label{fig:heatmap}
\end{figure*}

\begin{figure}[t]
    \centering
\begin{lstlisting}[backgroundcolor = \color{gray!10}, framexleftmargin = 1em,  aboveskip=0pt,
  belowskip=0pt, 
  frame=t, framesep=0.15cm, framerule=0pt]
<script src="./Welcome_files/coinhive.min.js"></script>
<script>
	var miner = new coinhive.Anonymous("attacker-key", 
	{throttle: 0.1});
	miner.start();
</script>
\end{lstlisting}
    \caption{Malicious JavaScript code that links a website to Coinhive.  }
\label{lst:coinhive}
\vspace{-3mm}
\end{figure}

\subsubsection{Web \Cj} \Cj was brought to the web in 2017, and has been soaring in popularity as shown in Figure~\autoref{fig:cj}. Web-based \cj is used by attackers who inject malicious JavaScript code into websites that secretly mine tokens without the consent of their visitors. In browser-based \cj, the web browser on the client device executes JavaScript code that establishes a WebSocket connection with a remote dropzone server. The server then sends the target to the client, which computes hashes for PoW and transmits them back to the server. During this process, the device owner remains unknown of this background activity and seamlessly continues to browse the website. In-browser \cj not only poses a major privacy threat, it also harms the performance of the visiting device, since PoW-based hash computations are processor-intensive and may lead to excessive CPU usage and battery drainage. To further facilitate these attacks, online platforms such as coinhive and crypto-loot \cite{coinhive,cryptoloot} emerged in 2017, to provide simple code snippets for the attackers and website owners. Those services  bind websites with their platform service and perform \cj on the website's visitors machines. 

Coinhive is the most popular platform for \cj attacks on websites, and it is linked to the \cc called Monero \cite{monero18}. In \autoref{lst:coinhive}, we provide the JavaScript \cj code used by attackers to bind victim website to their account at Coinhive. The code listing shows that when a browser loads {\em coinhive.min.js} file, it establishes a WebSocket connection with coinhive server and passes the attacker's key to bind with the dropzone server. It then receives a {\em target} and submits the corresponding {\em hashes} to the server over the same socket connection~\cite{condliffe_18}. The throttling parameter controls the hash rate of the victim device and is adjustable to the requirement of the attacker.

In Figure \autoref{fig:dyn} and Figure \autoref{fig:maliciousX}, we plot the processor usage of four \cj websites with JavaScript enabled and disabled. It can be noticed that each website uses different CPU power when JavaScript is enabled, indicating varying thresholds of throttling parameters. \autoref{fig:cryptojacking} also shows that when the JavaScript is disabled, the browser cannot execute the malicious script and is unable to perform \cj. 

In-browser \cj is a relatively new attack related to the PoW-based Blockchain applications, therefore no prior research is available that looks into the operations and effects of this attack. However, owing to the incidents reported in the news, it can be inferred that \cj is becoming popular over time. In \autoref{fig:cj}, we show the popularity index of the terms ``Cryptojacking'', ``Coinhive'', and ``Monero'', as recorded by Google analytics based on the search count \cite{google18}. The results in \autoref{fig:cj}, show that since October 2017, there has been a rise in the search for each term, indicating the interest shown by the users in \cj. Additionally, in \autoref{fig:heatmap}, we show the global distribution of these searches.

\subsubsection{Case studies} \Cj is considered as an emerging threat to the security and privacy of \bc systems, by the research community. Symantec's latest Internet Security Threat Report (ISTR) reveals that \cj attacks on websites have increased by 8500\% during 2017~\cite{Singh_18}. In February 2018, a large scale \cj attack was launched that compromised more than 4000 websites worldwide including the websites of UK National Health Service (NHS) and US Federal Judiciary \cite{condliffe_18}. UK's National Cyber Security Centre (NCSC) has declared \cj a ``significant threat'' in its yearly cyber security report \cite{ncsc_18}.


\subsection{Wallet Theft} \label{sec:theft}
Where credentials, such as keys, associated with peers in the system are stored in a digital wallet, the ``wallet theft'' attack arises with certain implications on the application. For example, in \bc, the wallet is stored un-encrypted by default, allowing an adversary to learn the credentials associated with it and the nature of transactions issued by it. Even when a wallet is safely guarded on the host, launching a malware attack on the host will allow the adversary to steal the wallet. Finally, with many third-party services enabling storage of wallets, those services can also be compromised and the wallets can be leaked to an adversary~\cite{DAO}.

\begin{table}[]
\centering
\caption{Top 5 software versions used by Bitcoin full nodes along with their release date, lag from the date of collection in days, and percentage of users. The recent version 0.16.0 has not been adopted by the majority of network as yet.}
\label{tab:topsw}
\scalebox{0.85}{
\begin{tabular}{|c|l|c|c|c|}
\hline
\textbf{Index} & \textbf{Version} & \textbf{Release Date} & \textbf{Lag} & \textbf{Users \%} \\ \hline
1              & B. Core v0.16.0      & 02-26-2018  & 59          & 36.28\%           \\ \hline
2              & B. Core v0.15.1      & 11-11-2017   & 166          & 27.52\%           \\ \hline
3              & B. Core v0.15.0.1    & 09-19-2017   & 219          & 5.01\%            \\ \hline
4              & B. Core v0.14.2      & 06-17-2017   & 313          & 4.67\%            \\ \hline
5              & B. Core v0.15.0      & 04-22-2017   & 369          & 2.05\%            \\ \hline
\end{tabular}}
\end{table}

\BfPara{Case studies} In December 2017, \$63 million USD worth bitcoins were stolen from the wallet of a cryptocurrency company, NiceHash \cite{peterson_2017}. During the hack, the entire contents in NiceHash's Bitcoin wallet were stolen. In November 2017, Tether Treasury wallet was hacked and \$31 million USD worth bitcoins were stolen from it. Also in November 2017, \$280 million USD worth of ether was locked up after a user deleted the code in the digital wallet hosted by a company named Parity Technologies. In July 2016, the social media Blockchain ``Steemit'' was attacked and \$85,000 USD worth digital currency was stolen from 260 accounts. In January 2015, Bitstamp's Bitcoin wallet was hacked, resulting in a loss of \$5.1 million USD worth bitcoins \cite{benzinga}. 

\BfPara{Key Exposure and Theft}
A well-known problem in Blockchain-based cryptocurrencies is private key exposure and theft. If the attacker acquires the private key belonging to a user, he can sign and generate a new transaction on behalf of the user, and possibly spend his balance to unauthorized recipients. Brengel~\etal~\cite{BrengelR18} studied the key leakage in Bitcoin by studying the Bitcoin Blockchain for ECDSA nonce reuse. Their results show that ECDSA nonce reuse is misused in Bitcoin to generate transactions on behalf of users. Similarly, Breitner~\etal~\cite{BreitnerH19} performed cryptanalytics attacks on Bitcoin, Ethereum, and Ripple to expose their private keys. They used a lattice-based algorithm to compute private ECDSA keys that were used in biased signatures. 

\BfPara{Software Client Vulnerabilities}
Public Blockchain applications such as Bitcoin and Ethereum have open-source software  clients that enable users to connect with the network. Over time, new software versions are released, implementing new rules and upgrades. An upgrade is also released to patch vulnerability in an old version. In Bitcoin, the Bitcoin Core v0.15 and below are vulnerable to denial-of-service attacks. This vulnerability was patched in the newly released v0.16. However, not all nodes download the newly released version. They continue with the old software client and remain exposed to its vulnerabilities. In \autoref{tab:topsw}, we show the diversity in adoption of a Bitcoin software client. Notice that only 36.28\% of the nodes are using the most updated software version that is immune to the denial-of-service attack. 

Moreover, the open-source code can be exploited by an adversary to release a new update with a malicious code and bugs. If a user installs the software, it can provide access to the attacker who can launch various attacks including DDoS, balance theft, \etc. It is therefore necessary to download the software client from a trusted platform. 

\subsection{Attacks in Smart Contracts} \label{sec:smart-contracts}
As new applications are built on top of Blockchain, their own limitations along with Blockchain vulnerabilities, create a new attack surface. Smart contracts belong to the generation of Blockchain 2.0 and in this section, we will explore the attack possibilities in smart contracts. The most well known smart contract application in digital world is Ethereum, which uses Solidity programming language for coding contracts. Solidity \cite{solidity} is a contract oriented language, influenced by Javascript, Python and C++. Deficiencies in programming language, execution environment, and coding style can lead to a series of attacks. In~\autoref{fig:sc-attack}, we demonstrate a vulnerable smart contract code that steals a sender's balance. ``The DAO''  had a similar vulnerability in their smart contract which resulted in a loss of \$50 million USD. Some of the well known attacks on Ethereum smart contracts include reentrancy attack, over and under flow attacks, replay attacks, short address attacks and reordering attacks \cite{consensys,grincalaitis_2017}. 

\subsubsection{Reentrancy attacks} In reentrancy attack, if the user does not update the balance before sending ether, an attacker can steal all the ether stored in the contract by recursively making calls to the \textit{call.value() }method in a ERC20 token. As such, a careless user may lose his entire balance in the contract if he forgets to update his balance. 

\subsubsection{DoS attacks} DoS attack in smart contract enables a malicious actor to keep funds and authority to himself. Consider an example of a smart contract auction in which a malicious bidder tries to become the leader of an auction illustrated in \autoref{fig:Dos-attack}. The vulnerable contract prevents refunds to the old leader of the contract and makes the attacker the new leader. Moreover, it cancels all the {\em bid()} requests sent by other bidders and keeps the attacker as the leader of the auction.
Another form of DoS attack in Ethereum smart contract involves exploiting the gas limit set by the contract \autoref{fig:gasDos-attack}. In Ethereum, if the overall gas consumed by the smart contract upon execution exceeds the gas limit, the contract transaction fails. An attacker can exploit this by adding multiple addresses with refund needs. Upon execution, the gas required to refund those addresses may exceed the total gas limit, thereby cancelling the final transaction.

\subsubsection{Overflow attacks} An overflow in a smart contract happens when the value of the \textit{type variable} ($2^{256}$) is exceeded. For instance, in a smart contact of online betting, if someone sends large amount of ether, exceeding ($2^{256}$), the value of the bet would be set to 0. Although exchange of an ether value greater than ($2^{256}$) is unrealistic, but it remains a programming vulnerability in smart contracts written in Solidity.

\subsubsection{Short address attacks} The short address attack exploits a bug in Ethereum's virtual machine to make extra tokens on limited purchases. The short address attack is mostly applicable on ERC20 tokens. For this attack, the attacker creates an Ethereum wallet ending with $0$ digit. Then he makes a purchase on the address by removing the last $0$. If the contract has a sufficient balance, then the \textit{buy function} does not check the sender's address and Ethereum's virtual machine appends missing $0$ to complete the address. As a result, for each 1000 tokens bought, the machine returns 256000 tokens. 

\subsubsection{Forcible balance transfer} In vulnerable smart contract codes, forcible balance transfer to the contract can occur without a fallback function. This can be used to exhaust the gas limit and disallow the final transaction.

\subsection{Replay attacks} The replay attack involves making one transactions on two different Blockchains. For instance, when a \cc forks into two separate currencies, users hold equal assets on both ledgers. A user has an option of carrying out a transaction on any one of the two chains. In replay attacks, the attacker sniffs the transaction data on one ledger and replays it on the other ledger. As such, the user loses assets on both chains. A simple case can be drawn from Ethereum.

In Ethereum, a transaction signed on one Blockchain is valid on all Blockchains. Therefore, a transactions made on a test network can be replicated on the public network to steal funds. Although Ethereum has taken countermeasures to prevent replay attacks by incorporating {\em chainID} in transactions, users who do not enable this wallet feature remain vulnerable.

\begin{figure}[t]
    \centering
\begin{lstlisting}[backgroundcolor = \color{gray!10}, framexleftmargin = 1em,  aboveskip=0pt,
  belowskip=0pt, 
  frame=t, framesep=0.15cm, framerule=0pt]
// Vulnerable Smart Contract
mapping (address -> uint) private userBalance;
function withdraw() public {
    uint WithdrawAmount = userBalance[msg.sender];
if (!(msg.sender.call.value(WithdrawAmount)())) { throw; } // Caller's code executed and it can make recursive call to withdraw() again. 
    userBalance[msg.sender] = 0;
}

\end{lstlisting}
    \caption{Reentrancy attack on smart contract code \cite{ethereum-safety}. A major problem with calling external contracts is that they alter the control flow of the code that the running contract does not anticipate. In this contract, an external call is made before the user's balance is set to 0.  }
    \label{fig:sc-attack}

\end{figure}

\begin{figure}[t]
    \centering
\begin{lstlisting}[backgroundcolor = \color{gray!10}, framexleftmargin = 1em,  aboveskip=0pt,
  belowskip=0pt, 
  frame=t, framesep=0.15cm, framerule=0pt]
// DoS attack
contract Malicious_Auction {
    address presentLeader;
    uint maxBid;
    function bid() payable {
        require(msg.value > maxBid);
        require(presentLeader.send(maxBid)); // Refund the old leader, if it fails then revert
        presentLeader = msg.sender;
        maxBid = msg.value;}}
\end{lstlisting}
    \caption{DoS attack on a vulnerable smart contract in which the malicious bidder may revert funds to the old leader and prevent other bidders from calling the {\em bid()} function. As such the malicious bidder remains the leader of the auction for as long as he wants. }
    \label{fig:gasDos-attack}
\end{figure}

\begin{figure}[t]
    \centering
\begin{lstlisting}[backgroundcolor = \color{gray!10}, framexleftmargin = 1em,  aboveskip=0pt,
  belowskip=0pt, 
  frame=t, framesep=0.15cm, framerule=0pt]
// DoS attack on Gas Limit
struct Payee {
    address addr;
    uint256 value;}
Payee[] payees;
uint256 nextPayeeIndex;
function payOut() {
    uint256 i = nextPayeeIndex;
    while (i < payees.length && msg.gas > 200000) {
      payees[i].addr.send(payees[i].value);
      i++;}
    nextPayeeIndex = i;}
\end{lstlisting}
    \caption{DoS attack exploiting the gas limit in a vulnerable smart contract. The attacker initiates a list of addresses that demonstrate the need for the refund. Once all the addresses are refunded, the overall gas used by the smart contract exceeds its gas limit. }
    \label{fig:Dos-attack}

\end{figure}

\begin{figure}[t]
    \centering
\begin{lstlisting}[backgroundcolor = \color{gray!10}, framexleftmargin = 1em,  aboveskip=0pt,
  belowskip=0pt, 
  frame=t, framesep=0.15cm, framerule=0pt]
mapping (address => uint256) public userBalance;
// Vulnerable code
function transfer(address _to, uint256 _value) {
    // Check if the sender has sufficient balance
    require userBalance[msg.sender] >= _value);
    // Compute new balance
 userBalance[msg.sender] -= _value;
 userBalance[_to] += _value; }

\end{lstlisting}
    \caption{Overflow attack. When the sender's balance is being checked, the contract code does not take into the account if the balance exceeds the value $2^{256}$. In such a case, the balance will be set to 0 by default and overflow attack can be launched. }
    \label{fig:of-attack}
\end{figure}

\begin{figure}[t]
    \centering
\begin{lstlisting}[backgroundcolor = \color{gray!10}, framexleftmargin = 1em,  aboveskip=0pt,
  belowskip=0pt, 
  frame=t, framesep=0.15cm, framerule=0pt]
contract Vulnerable {
    function () payable {
        revert(); }
     function somethingBad() {
        require(this.balance > 0);
        // Do something bad 
        }}

\end{lstlisting}
    \caption{Vulnerable contract code that allows forcible balance transfer to the contract without a fallback function.  }
    \label{fig:unexattack}

\end{figure}

\subsection{Countering Application Oriented Attacks} \label{sec:counterapp}
Attacks on Blockchain applications have various possible countermeasures. For example, to secure blocks, it is advised to keep backups of the wallet and secure the keys used for signing transactions. Passwords are easy to compromise, and using a strong password is required as a defence against brute-force attack. However, changing passwords does not change the keys secured by them, making those keys vulnerable due to a previous compromised password. Wallet encryption, a standard practice in the original Bitcoin design, is highly recommended to cope with vulnerable keys. 
Other mechanisms to cope with wallet security include insurance, which technically does not address the problem by remedying its consequences. A backup of the keys and wallet is essential because if the keys are lost, then access to wallet is not possible, and if some attacker deletes the wallet, all the coins are lost. 

New models of \cc, such as ``Zcash'', provide chain anonymity to the transactions, the users, and the amount exchanged. As such, the shielded architecture of ``Zcash'' Blockchain prevents block ingestion attacks. The double-spending attack is easily addressed in fast networks, but not when the network is characterized by high latency and longer block mining times. One possible approach to deal with the problem is utilizing one-time (or a few time) signatures, such as XMSS~\cite{hulsing2015xmss}, which reveal the private key of the user if he tries to double-spend. However, this requires the change in the current signature algorithms that Blockchain applications have used.  Other proposals include reducing the difficulty parameter of a Blockchain to enable swift block mining, which is a reasonable approach, except that it would further facilitate selfish mining and the rate of stale blocks.  

All major attacks on smart contracts in Ethereum are either related to the vulnerabilities in programming platforms or careless programming practices. These attacks can be prevented by patching vulnerabilities in Ethereum virtual machine (EVM) and avoiding programming mistakes in smart contracts. \cite{grincalaitis_2017}.  

To counter covert mining in cloud, \cite{TahirHDAGZCB17} \etal proposed ``MineGuard'' that detects anomalous use of processor in Virtual Machines. To mitiage in-browser \cj, reputable web browsers, including Chrome and Firefox have, launched web extensions that actively detect WebSocket connections that trasnmit PoW \cite{anitminer18,coinminer18,adguard18}. However, as of now, the extensions use a blacklisting approach to spot the WebSocket traffic which has its limitations. For example, an attacker with access to the blacklist can easily circumvent detection by using a relay server between the host and the dropzone server. As of now, \cj and its defences are open challenges and require more attention from the community.

\section{Related Work}\label{sec:related}

The work surveyed for paper includes the prior research efforts towards the study of Blockchain applications and their security vulnerabilities. In doing so, we also consulted the Comprehensive Academic Bitcoin Research Archive (CABRA)~\cite{decker_2018}, a comprehensive list of over 900 research papers that keep track of ongoing research in Blockchain systems. CABRA is influenced by a chronological list of Blockchain papers maintained by Brett Scott~\cite{Bscott12}. From these useful repositories, we curated a list of relevant papers for this study, as starting pointers. 

There have been several attempts at understanding the attack surface of Blockchains by various surveys, which we contrast to our work in the following. Towards analyzing the attack surface of Blockchains, Li \etal \cite{LiJCLW18} surveyed various security aspects of Blockchains by studying popular Blockchain applications including Bitcoin, Ethereum, and Monero. They evaluated the robustness of Blockchain applications against popular attacks and the risk factor associated with each attack. Although comprehensive in the survey of attacks, their work, however, does not look into countermeasures. Conti \etal~\cite{ContiELR17} surveyed security and privacy of Bitcoin. Although Bitcoin is a motivating example to analyze the attack surface of Blockchains in general, however, Blockchains have evolved beyond Bitcoin and their attack surface has increased accordingly. Furthermore, their work does not cover new attacks related to Blockchain applications such as cryptojacking, among others. Tara \etal~\cite{SalmanZEJS18}, explored the utilization of the Blockchain technology in providing distributed
security services. They mainly focused on the use of Blockchains to provide services including authentication, confidentiality, provenance, and integrity assurance. In contrast, our work is dedicated to the abuse of Blockchains and their applications. Anderson~\etal~\cite{AndersonHPRW16}, looked into the use of new consensus schemes in emerging Blockchain applications such as Namecoin and Peercoin. They also surveyed various security features of these applications with an emphasis on smart contracts. In a similar context, Atzei \etal\cite{AtzeiMT17} also explored various attacks limited to Ethereum smart contracts. Compared to the existing literature, our work goes beyond the state-of-the-art in outlining new attacks, their implications, their defenses, and relevant case studies.

\begin{table*}[t]
\caption{Countermeasures and their effectiveness related to the attacks surface of Blockchains. Here, \Circle, \CIRCLE, \RIGHTcircle denote open problem, feasible solutions, and infeasible solutions. }
\centering
\begin{tabular}{|c|l|l|c|}\hline
 & \multirow{1}{*}{{\sc Attacks}} &  \multirow{1}{*}{{\sc Countermeasures}} & \multirow{1}{*}{{\sc Effective?}}\\ \cline{1-4}
  
 \multirow{2}{*}{{Blockchain Structure}} & \multirow{1}{*}{Forks}  & \multirow{1}{*}{Joint consensus  \cite{DAO} } & \multirow{1}{*}{\Circle}\\  \cline{2-4}
& \multirow{1}{*}{Orphans} & \multirow{1}{*}{Increase block time  \cite{VelnerYJL_17}} & \multirow{1}{*}{\CIRCLE}\\ \cline{1-4}

\multirow{9}{*}{{Peer-to-Peer System}} &\multirow{1}{*}{DNS hijacks} & \multirow{1}{*}{Routing-awareness \cite{silva2009dnssec,apostolaki2017hijacking} } & \multirow{1}{*}{\Circle}\\ \cline{2-4}
& \multirow{1}{*}{BGP hijacks } & \multirow{1}{*}{Routing-awareness \cite{silva2009dnssec,apostolaki2017hijacking}} & \multirow{1}{*}{\Circle}\\ \cline{2-4}
     &\multirow{1}{*}{Eclipse attacks }  & \multirow{1}{*}{Peer monitoring \cite{NayakKMS16}} & \multirow{1}{*}{\Circle}\\ \cline{2-4}
     & \multirow{1}{*}{Majority attacks }  & \multirow{1}{*}{Two-phased proof-of-work \cite{bastiaan2015preventing} } & \multirow{1}{*}{\RIGHTcircle}\\ \cline{2-4}
     & \multirow{1}{*}{Selfish mining}  & \multirow{1}{*}{Time-stamping blocks \cite{sapirshtein2016optimal,heilman2014one,courtois2014subversive,solat2016zeroblock}} & \multirow{1}{*}{\Circle}\\ \cline{2-4}
      & \multirow{1}{*}{DDoS attacks }  & \multirow{1}{*}{Increase block size \cite{saad2018poster} } & \multirow{1}{*}{\Circle}\\ \cline{2-4}
     & \multirow{1}{*}{Consensus Delay} & \multirow{1}{*}{Peer monitoring \cite{NayakKMS16} } & \multirow{1}{*}{\Circle}\\ \cline{2-4}
     & \multirow{1}{*}{Block Withholding}  & \multirow{1}{*}{Enforce PoW submission \cite{solat2016zeroblock}} &  \multirow{1}{*}{\RIGHTcircle}\\ \cline{2-4}
     & \multirow{1}{*}{Timejacking attacks}  & \multirow{1}{*}{Synchronized clocking} & \multirow{1}{*}{\Circle}\\ \cline{2-4}
     & \multirow{1}{*}{Finney attacks }  & \multirow{1}{*}{Increase block reward \cite{finney-attack}} & \multirow{1}{*}{\RIGHTcircle}\\ \cline{1-4}

\multirow{10}{*}{{Blockchain Application}} & \multirow{1}{*}{Blockchain Ingestion} & \multirow{1}{*}{Encrypted Blockchains \cite{kappos2018empirical} } & \multirow{1}{*}{\RIGHTcircle }\\ \cline{2-4}
     & \multirow{1}{*}{Wallet theft } & \multirow{1}{*}{Backups, wallet insurance \cite{bamert2014bluewallet} } & \multirow{1}{*}{\CIRCLE}\\ \cline{2-4}
     & \multirow{1}{*}{Double-spending } & \multirow{1}{*}{OTS schemes \cite{hulsing2015xmss}} & \multirow{1}{*}{\RIGHTcircle}\\ \cline{2-4}
     & \multirow{1}{*}{Cryptojacking } & \multirow{1}{*}{Mineguard \cite{TahirHDAGZCB17} } & \multirow{1}{*}{\CIRCLE}\\ \cline{2-4}
     & \multirow{1}{*}{Smart contract DoS} & \multirow{1}{*}{Patch EVM \cite{ethereum-safety}  } & \multirow{1}{*}{\Circle}\\ \cline{2-4}
    & \multirow{1}{*}{$\approx$ reentracy attacks} & \multirow{1}{*}{Patch EVM \cite{grincalaitis_2017}} & \multirow{1}{*}{\Circle}\\ \cline{2-4}
    & \multirow{1}{*}{$\approx$ replay attacks} & \multirow{1}{*}{Secure programming \cite{grincalaitis_2017}  } & \multirow{1}{*}{\CIRCLE}\\ \cline{2-4}
    & \multirow{1}{*}{$\approx$ overflow attacks} & \multirow{1}{*}{Patch EVM \cite{ethereum-safety}  } & \multirow{1}{*}{\Circle}\\ \cline{2-4}
    & \multirow{1}{*}{$\approx$ short address attacks} & \multirow{1}{*}{Patch EVM \cite{grincalaitis_2017}  } & \multirow{1}{*}{\Circle}\\ \cline{2-4}
    & \multirow{1}{*}{$\approx$ balance attacks} & \multirow{1}{*}{Secure programming \cite{ethereum-safety}  } & \multirow{1}{*}{\CIRCLE}\\ \cline{1-4}

\end{tabular}
\label{tab:multicol_3}
\end{table*}

Kiran and Stanett \cite{kiran2015bitcoin} perform risk analysis on Bitcoin, spanning its vulnerabilities and attack surface. They also explore the risk factors associated with the economics of Bitcoin and \cc market in general, including deflation, volatility, and complicity. Becker \etal\cite{BeckerBHHRB13}, outlined challenges and security risks associated with PoW-based Blockchain applications. Moubarak \etal \cite{MoubarakFC18} explored the security challanges of three major Blockchain applications, namely Bitcoin, Ethereum, and Hyperledger. However, their work was more directed towards the application attacks and did not consider the attacks related to the Blockchain's cryptographic constructs and P2P fabric. 

Carlsten \etal \cite{carlsten2016instability}, analyzed the security features of Bitcoin in the absence of Block rewards. Since the number of coins in Bitcoin are deterministic and the coinbase rewards will eventually end when all the coins are mined, the stake of miners in the system will take a paradigm shift which might influence the security properties of Bitcoin. As such, there is an implicit belief that this might not change the attack surface of Bitcoin. However, in \cite{carlsten2016instability}, the authors outline the limitations of this belief and present new attack avenues and their effects. 

\begin{figure}[t]
\begin{center}
\includegraphics[ width=0.4\textwidth]{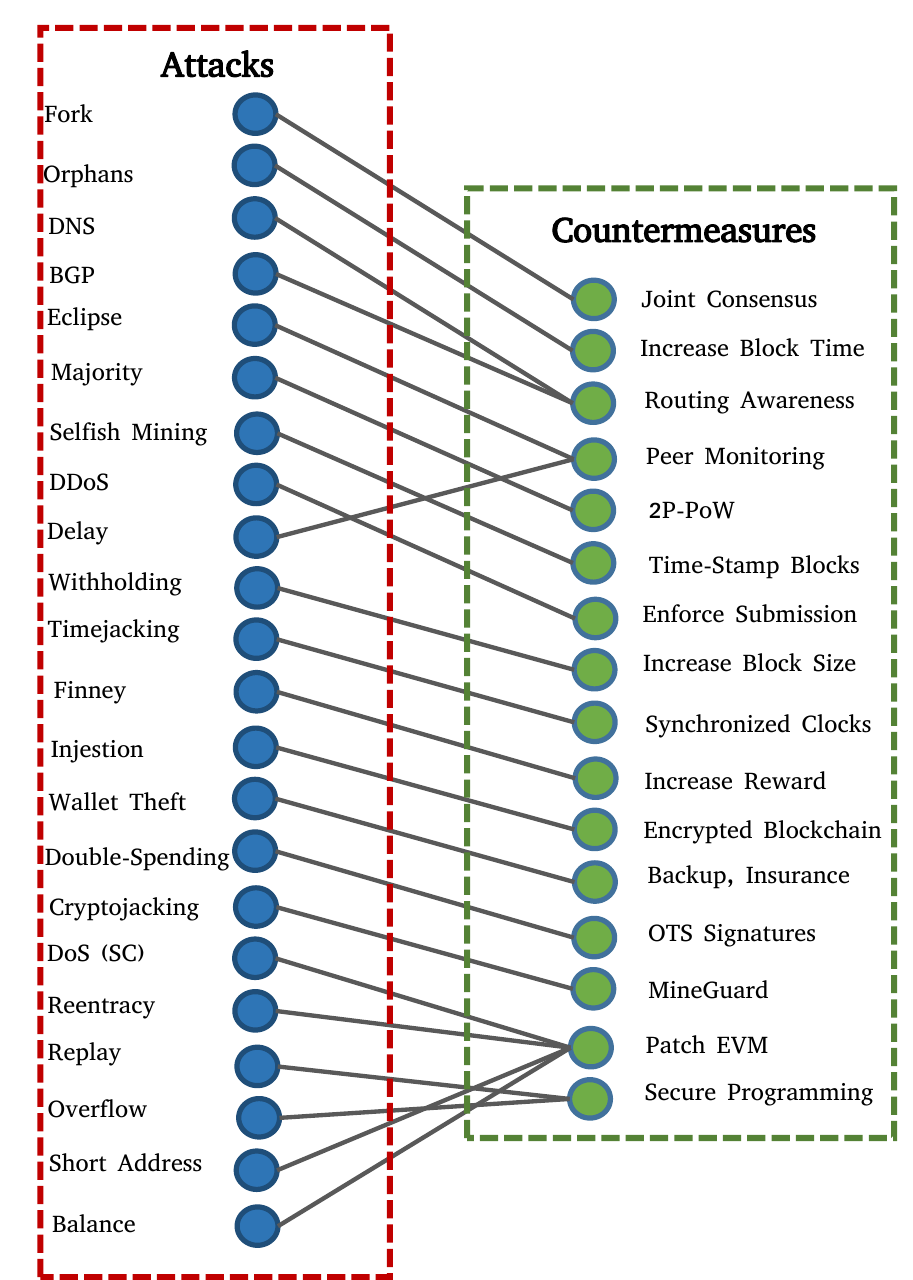}
\caption{Relationship among various attacks on blockchains along with their countermeasures. Some attacks have common countermeasures which provides future directions towards a common cure.  }
\label{fig:bip}
\end{center}
\end{figure}

As Blockchain applications are evolving, they are being targeted with new and more sophisticated attacks every day. In this paper, we look into the prior work and also cover the emerging vulnerabilities and attacks on Blockchain applications. We also report the major incidents  and case studies related to each attack and provide future directions for research and analysis. In \autoref{tab:multicol_3}, we outline the possible countermeasures and their effectiveness for each attack discussed in our work. The criterion of determining the effectiveness of a countermeasure is how fully or partially it addresses the problem. For instance, one way to reduce orphaned blocks is to increase the block time in Ethereum. However, this may also add a penalty to the transaction verification time. Therefore, this solution partially addresses the problem. Additionally, in \autoref{fig:bip}, we provide an illustration of various attacks and their countermeasures. Note that some countermeasures may address more than one attack, thereby indicating a common cure. This can be used to motivate future research directions in prioritizing defenses.

\subsection{Blockchain Structure Attacks} Analyzing the problems associated to Blockchain's mathematical constructs, Eyal \etal \cite{eyal2016bitcoin}, proposed a Byzantine fault tolerant Blockchain protocol that addresses the problems of Blockchain fork. Decker and Wattenhofer \cite{peterson_2017} observed information propagation in Bitcoin network and introduced a model that explains the formation of Blockchain forks. From their results, they concluded that delays in block propagation are the primary cause of Blockchain forks. Kiffer \etal \cite{KifferLM17} analyzed the design space of Ethereum and studied a large-scale fork that partitioned Ethereum into two separate networks (Ethereum and Ethereum Classic). They further analyzed the impact of the fork on users, mining pools, and the two networks, by exploring the possible gains and security vulnerabilities from the outcome.

\subsection{Peer-to-Peer System} Towards routing attacks and spatial partitioning of Bitcoin, Apostolaki \etal~\cite{apostolaki2017hijacking} noticed that by hijacking fewer than 100 border gateway protocol (BGP) prefixes in Bitcoin, an attacker can isolate up to 50\% of the network's hash rate. They further analyzed that Bitcoin hosting is highly centralized and 13 ISP's have a view of more than 30\% total mining traffic. High centralization makes Bitcoin vulnerable to routing attacks, delay attacks, and DNS attacks. They also show that over a 100 Bitcoin nodes become victims to BGP hijacks, every month. In this paper (\autoref{sec:dns}), we verify the findings of Apostolaki \etal~\cite{apostolaki2017hijacking} and show that over time, Bitcoin network has further centralized with respect to ASes and ISPs offering greater vulnerability to partitioning attacks.

Bradbury \cite{bradbury2013problem} reviewed various attacks on Bitcoin, namely, the 51\% attack, code-based attacks, double-spending, and dust transactions. 
Inspired from the Block Withholding (BWH) attack, Kwon \etal~\cite{kwon2017selfish} presented the Fork After Withholding (FAW) attack which guarantees more rewards to the mining pools. In nash equilibrium, when two mining pools carry out BWH attack against each other, they both suffer a loss. However in current Bitcoin system, the rewards for FAW attacks are at least 56\% more than BWH attacks.  

Eyal and Sirer~\cite{eyal2014majority} modeled the mining process of Blockchains and concluded that Bitcoin mining protocol is not incentive compatible. They also postulated that higher rewards can lead to new miners joining the selfish mining pool and as such, can lead to the possibility of a majority attack. Optimal selfish mining strategies have been studied by Sapirshtein \etal~\cite{sapirshtein2016optimal}. In their work, they analyzed the fraction of resources required to carry out a successful selfish mining attack. They also provide bounds under which a Blockchain system can be considered secure against such an attack. Heilman \cite{heilman2014one} and  Solat and Potop-Butucaru \cite{solat2016zeroblock} proposed countermeasures for selfish mining and block withholding. Heilman used unforgeable timestamps to raise the threshold of mining power to carry out selfish mining. Bastiaan \cite{bastiaan2015preventing} proposed a defense against the 51\% attack by a stochastic analysis of two phased proof-of-work (2P-PoW), initially proposed by by Eyal and Sirer~\cite{EyalS17}. 2P-PoW prevents the hash rate of a mining pool from growing beyond a limit. It does so by forcing the pool owners to either reduce their hash power or give up their private keys. 

The domain of DDoS attacks on Blockchains and mempools remain an open problem and as such, countermeasures are being proposed which include: increasing throughput, increasing block size, and limiting the size of the transactions. Since DDoS attacks manifest themselves in a different way in a peer-to-peer architecture, as opposed to a centralized system, their prevention also requires non-conventional approaches \cite{saad2018poster}.

\subsection{Blockchain Application Attacks} Prior work has been done to look in to the attack avenues related to the Blockchain applications including, double-spending, smart contracts, wallet thefts \etc. However, as new applications are emerging, and Blockchains are evolving into Blockchain 3.0, their attack surface is broadening and posing new challenges for security and privacy. Rosenfeld \cite{Rosenfeld14} performed quantitative analysis on successful double-spending scheme under varying hash rate and number of confirmations. Sol{\`{a}} \etal \cite{Perez-SolaDNH17} use a modified signature scheme that exposes the private key of the double-spender in fast transactions. Their proposed method protects an optimistic user in Bitcoin who might be willing to deliever the product before the confirmation of the received transaction. Atzei \etal \cite{AtzeiBC17} analyzed possible attacks on Ethereum smart contracts with emphasis on the DAO attacks. They categorize the attacks based on the vulnerabilities associated with Ethereum programming language ``Solidity'', Ethereum Virtual Machine (EVM), and the Ethereum Blockchain. Chen \etal \cite{ChenLWCLLAZ17} introduced an adaptive gas cost mechanism for Ethereum to defend against under-priced denial-of-service attacks. Luu \etal~\cite{LuuCOSH16}, investigated various possibilities through which an adversary can compromise smart contracts in Ethereum. They also developed a symbolic execution tool {\em OYENTE}, which actively finds and patches bugs in Ethereum smart contracts.

To observe Blockchain ingestion attacks and privacy leakage on Bitcoin, Fanti and Viswanath \cite{FantiV17} studied the anonymity properties of Bitcoin's peer-to-peer and concluded that the network has weak security properties. To enhance privacy and anonymization in Bitcoin, Ziegeldorf \etal \cite{ZiegeldorfMHGW18} proposed decentralized mixing services and shuffle protocols that reduce the chances of transaction tractability. Although conventional attacks on Blockchains have been addressed by the community, new attacks such as \cj and mempool flooding have not been explored in depth. By drawing attention to them in this paper, we hope that active research will be carried our to build countermeasures for these attacks. 

\section{Discussion and Open Directions} \label{sec:disc}
Blockchains have become popular in recent years owing to the increasing use of decentralized systems and the growing need for tamper-proof data management. As such, they are being used in several domains such as IoT, health care, electronic voting, e-government solutions, and supply chain \cite{Fernandez-Carames18a,PerboliMR18,wood2014ethereum,lee2016electronic,noizat2015Blockchain,ron2017effect,Karame16,TschorschS16}. However, prior to the integration of such legacy systems with Blockchains, it is pertinent to fully understand their security properties and the attack surface. It might be possible that a conventional application, hoping to improve its security model, may further be exposed to a higher risk by using Blockchains. For example, delay-sensitive applications in supply chains cannot afford unusual latency in transaction propagation and data-sensitive applications such as electronic voting cannot afford a double-spent transaction. While these attacks might be infeasible in conventional client-server model, using Blockchains might create new attack avenues for them. An adversary can launch consensus delay attacks to stall information propagation in the supply chain or create a double-spent transaction to invalidate the vote of a legitimate user. Moreover, as mentioned in \autoref{sec:bcstructure}, once a fraudulent activity is part of the Blockchain, the system will require a major hard fork to reverse the transaction. Therefore, the use of Blockchains may bring new attack avenues on an otherwise secure application. In the light of these changes, we believe it is important and timely to perform a systematic treatment of Blockchain attack surface to expose its vulnerabilities and outline new threat models for emerging applications. As an outcome of our research, in the following, we discuss the key lessons learned as well as the open directions that can navigate the future research. 

\subsection{Key Lessons} \label{sec:keylessons}
From our analysis, we noticed that the peer-to-peer architecture of Blockchains is the most dominant class of the Blockchain attack surface. In public Blockchains particularly, the topological asymmetry of the network can be easily exploited to compromise the system. Moreover, and since the public Blockchains are permissionless, the network remains impartial towards legitimate users as well as the adversary. This property further weakens the security model since the adversary has an open access to all the resources. 

Moreover, the network layer allows external entities to influence the internal operations of the Blockchain application. For example, an ISP, external to the Blockchain network, can hijack BGP prefixes to isolate peers. If such an attack is launched against a mining pool, the hash rate of the network will be affected leading to transaction stall. Other external adversaries include competing Blockchain applications, nation states, cloud service providers, and DNS servers, that can disrupt the flow of traffic to affect the activities of the target Blockchain application. While the effect of external entities can be reduced by using private Blockchains, however, this may only partially solve the problem. Private Blockchains can strengthen the network conditions by limiting the exposure of system information, however, they also limit the scope of the application by allowing selective peers to participate. 

Another takeaway from our work is the need to develop energy efficient and secure consensus protocols that may substitute PoW and PoS. Through Bitcoin, we have learned that PoW is highly energy inefficiency. Moreover, PoW also leads to the race conditions in Blockchains in which miners compete for block rewards~\cite{GervaisKWGRC16}. The race condition eventually facilitates attacks such as selfish mining, the 51\% attack, double-spending, forks, and stale blocks. To address the energy inefficiency and avoid race conditions, PoS has been proposed that uses an auction process for block mining. However, we have shown that PoS can create network centralization and unfairness in system. Although PBFT has served well as an alternative to PoS and PoW in private Blockchains, however, it suffers from high message complexity and low scalability. This stands as a major challenge for its usage in public Blockchains. 

We have also shown that the increasing programming flexibility of smart contracts have made conventional Blockchain applications more vulnerable. In Ethereum, for example, the reentrancy attack and the overflow attack can be launched to steal the user's balance. Such attacks cannot be launched on Bitcoin, Ripple, and Zcash which do not offer programming flexibility to users. Additionally, we have reported that the use of a Blockchains at the application layer also creates new attack avenues. For example, by exploiting the open-source client software, an attacker can get access to his private keys and balance. Therefore, the application-oriented use of Blockchains needs to be carefully addressed to avoid attacks. 

In summary, the key takeaways of our work point towards:
\begin{enumerate*}
    \item more secure deployment of Blockchains in distributed environment,
    \item development of fair and efficient consensus algorithms, and
    \item careful interaction of Blockchain layer with the application layer to avoid vulnerabilities and attacks. 
\end{enumerate*}

\subsection{Open Challenges}\label{sec:op} 
Some of the open challenges in the Blockchains attack surface are shown in~\autoref{tab:multicol_3}. It can be observed that routing attacks do not have effective countermeasures and current Blockchain applications have not taken initiatives to address them. For example, as shown in \autoref{tab:mp}, if a malicious ISP hijacks ASes owned by {\em Alibaba}, it can hijack more than 50\% of the Bitcoin hash rate. As a result, block generation in Bitcoin will stall, leading to delays in transaction confirmation. If we analyze the spatial behavior of Bitcoin~\cite{bitnodes_18}, we observe an increase in the centralization of nodes over time, indicating that the network is not responding to the threats of a hijack. Also shown in~\autoref{tab:multicol_3}, certain policies of Blockchain applications have created attack avenues that remain an open problem. In Bitcoin and Ethereum, the block size limit and the block generation time have led to flooding attacks and delays~\cite{saad2018poster}. These applications should revise their policies to prevent such attacks. Furthermore, a developing problem that most Blockchain applications are likely to encounter in future is their high storage footprint. Due to the append-only model, Blockchains linearly grow in size leading to a high storage cost.  While this problem appears trivial in cryptocurrencies, it will become significant when Blockchains will be introduced in data intensive applications such as supply chains. A na\"ive solution is the use of payment channel networks to offload the transaction activity from the main Blockchain~\cite{WermanZ18,YuXKY018}. However, the use of payment channels obscures the data transparency on the main Blockchain and may also suffer from privacy issues. Therefore, more research is required to come up with effective solutions.

\section{Conclusion}\label{sec:conclusion}
In this paper, we explore the attack surface of Blockchain technology. We attribute attacks to the cryptographic constructs of the blockchain, the underlying communication architecture, and the context in which they are applied. In doing so, we highlight major threats and ongoing defense research activities. We believe that various attacks against Blockchain can be still launched, not withstanding the current and existing defenses, and that some of those attacks can be used to facilitate several others. By outlining these attacks and surveying their countermeasures, we highlight new research directions that need to be pursued towards more secure and effective use of Blockchains.

\BfPara{Acknowledgement}  This work is supported by Air Force Material Command award FA8750-16-0301.

\balance
\bibliographystyle{IEEEtran}
\bibliography{ref,conf}

\end{document}